\documentclass[11pt]{elsarticle}
\usepackage[top=2.0cm,bottom=2.5cm,left=2cm,right=2cm]{geometry}
\usepackage[utf8x]{inputenc}
\usepackage{helvet}
\usepackage{booktabs,nicefrac,microtype,natbib}
\usepackage{lineno,amsmath,color,soul,amsfonts}
\usepackage{float,graphicx,multirow,blindtext}
\usepackage[flushleft]{threeparttable}
\usepackage[outdir=./]{epsfig}
\modulolinenumbers[1]  
\usepackage{xfrac,bigints,enumitem,rotating}
\usepackage{relsize,longtable,mathtools,mathrsfs}
\usepackage[font=normalsize,labelfont={bf}]{caption}
\usepackage{siunitx}     
\usepackage{cancel}

\usepackage[colorinlistoftodos]{todonotes}
\usepackage{rotating}
\usepackage{pdflscape}
\usepackage{afterpage}
\usepackage{algpseudocode}
\usepackage{tikz}

\usepackage{esvect}

\usepackage{mathtools}

\usepackage[linesnumbered,ruled,longend]{algorithm2e}

\usepackage{esint}

\usepackage{tabularx, booktabs, threeparttable}

\usepackage{upgreek}  

\makeatletter
\def\ps@pprintTitle{%
  \let\@oddhead\@empty
  \let\@evenhead\@empty
  \def\@oddfoot{\reset@font\hfil\thepage\hfil}
  \let\@evenfoot\@oddfoot
}
\makeatother

\usepackage{hyperref}
\hypersetup
    {
    colorlinks=true,
    linkcolor=blue,
    filecolor=magenta,     
    urlcolor=cyan
    }
\hypersetup{pdfauthor={Name}}

\allowdisplaybreaks

\usepackage{pgfplots}

\DeclareMathOperator{\erf}{erf}

\begin{document}

\begin{frontmatter}

\title{Multiscale Dynamics of Roughness-Driven Flow in Soft Interfaces}

\author[mymainaddress]{Qian Wang}
\author[mymainaddress]{Suhaib Ardah\corref{mycorrespondingauthor}}
\author[mymainaddress]{Tom Reddyhoff}
\author[mymainaddress]{Daniele Dini}
\cortext[mycorrespondingauthor]{Corresponding author: s.ardah19@imperial.ac.uk}
\address[mymainaddress]{Department of Mechanical Engineering, Imperial College London, London, SW7 2AZ, UK}

\begin{abstract}
Soft lubricated contacts exhibit complex interfacial behaviours governed by the coupled effects of multiscale surface roughness and non-linear fluid-solid interactions. Accurately capturing this interplay across thin-film flows is challenging due to the strong synergy between contact mechanics and hydrodynamic flow, spanning over various spatiotemporal scales. Here, we develop a rigorous computational framework to simulate the frictional behaviour of soft lubricated interfaces; its modularity and the use of optimal solvers provides solutions for realistic configurations in lubrication regimes ranging from direct solid contact to complete fluid separation. Surface roughness is described via Persson’s statistical theory as well as a deterministic Conjugate Gradient with Fast Fourier Transform (CG-FFT) approach, while limitations associated with classical half-space models are addressed by developing the Reduced Stiffness Method (RSM) to rigorously model pressure-induced surface responses. The integrated framework captures the full evolution of frictional behaviour, validated against experiments on rough elastomer–glass interfaces, revealing how surface roughness and material compliance together drive the transition from solid contact to fluid-mediated sliding. The developed approach establishes a robust and versatile simulation tool for analysing a plethora of soft interfacial systems shaped by fluid–solid interactions, with potential applications including but not limited to biomechanics, soft robotics and microfluidic systems.
\end{abstract}


\end{frontmatter}


\section{Introduction}
Lubricated compliant interfaces are ubiquitous in nature and technology, underpinning systems as diverse as contact lenses \cite{dowson2009tribological}, biomedical implants \cite{licup2015stress}, tyres \cite{roberts1983friction} and smart adaptive materials \cite{rus2015design}. What unites these seemingly diverse systems is the presence of deformable materials that interact through a thin lubricating film. Unlike rigid contacting interfaces, compliant systems are characterised by large elastic deformations, expanded contact areas and complex fluid–solid interactions, making them functionally versatile but also mechanically intricate. In many engineering and biological applications, these interfaces operate in mixed or boundary lubrication regimes, where surface roughness becomes comparable to the lubricant film thickness \cite{Ardah2025}. As a result, the load applied on the contacting surfaces is shared between the entraining fluid and micro-asperity contacts, governed by multiscale deformation and topographical features. Accurately simulating the coupling between roughness, fluid flow and material compliance is essential for predicting friction, wear and overall system performance \cite{Vakis2018Sep}. This requires robust modelling frameworks capable of resolving fluid-solid interactions (FSIs) that occur across a broad spectrum of operating conditions.

Theoretical and computational modelling of compliant rough contacts has received significant attention, driven by the challenges associated with characterising lubricant film thickness and asperity-level interactions experimentally \cite{Muser2017Aug, pradhan2025surface}, especially under mixed and boundary lubrication regimes. Existing numerical strategies that tackle interfacial rough contacts can broadly be classified into three categories. The first class comprises stochastic models, which describe surface roughness via a limited set of statistical parameters that represent its random features \cite{thomas1998rough, whitehouse2002surfaces, Taylor2022May}. Among these, the widely adopted Patir and Cheng model \cite{Patir1978Jan, Patir1979Apr} remains a foundational approach for capturing roughness effects in lubricated interfaces, as it provides an efficient means of estimating film thickness and flow factors based on roughness statistics and surface orientation. However, its formulation captures only the global influence of surface topography and relies heavily on idealised statistical descriptors which can compromise accuracy, particularly when addressing real, measured surfaces. Deterministic models overcome this limitation by utilising actual surface profiles obtained through experimental measurements \cite{Jiang1999Jul, hu2001computer}, allowing for a more detailed resolution of asperity-level deformations and local contact pressures. Despite being more physically accurate, such approaches typically require fine spatial discretisations and become computationally demanding, especially when coupled with the hydrodynamic flow equations in a fully resolved FSI framework \cite{Ardah2025, Vakis2018Sep}. As such, they face a trade-off between physical accuracy and computational feasibility, thus limiting their practical application in multiscale or time-dependent analyses.

In order to bridge this gap, two-scale simulation frameworks have emerged as a powerful approach for modelling lubricated rough contacts, coupling macroscopic hydrodynamic flow with microscopic asperity interactions to achieve high predictive accuracy while avoiding the substantial computational expense of fully resolved fine-scale simulations \cite{persson2009transition, scaraggi2011lubrication, masjedi2014theoretical}. These approaches typically solve the hydrodynamic problem at the continuum scale while embedding reduced-order models that represent the influence of surface topography on local contact and fluid-flow behaviour. At the macroscopic level, hydrodynamic flow is commonly described using the classical Reynolds equation \cite{persson2009transition}, or its modified variants such as the Patir and Cheng model which introduce flow factors that account for anisotropy and pressure variations induced by surface roughness \cite{masjedi2014theoretical, sahlin2010mixed}. At the microscale, two main approaches have been widely adopted to model solid-solid interactions. The Greenwood and Williamson (GW) model \cite{masjedi2014theoretical, masjedi2015effect} idealises asperities as spherical summits, providing statistical estimates of contact load and area, whereas Persson’s theory \cite{scaraggi2011lubrication} captures the fractal nature of rough surfaces, offering analytical expressions for interfacial separation and real contact area. Although both models offer analytical convenience and computational efficiency, they fundamentally rely on simplified statistical representations of roughness that average out the spatial organisation of asperities. Comparative studies have shown that while GW and Persson-type formulations can reproduce broad contact trends \cite{carbone2008asperity, lorenz2010average}, their accuracy is highly sensitive to surface resolution and sampling window, making them less reliable for anisotropic or multi-scale topographies encountered in real systems. Moreover, these statistical models neglect local interactions between neighbouring asperities and fail to capture the non-linear coupling between roughness-induced pressure variations and material compliance. Consequently, despite their efficiency, existing two-scale frameworks remain limited in their ability to predict the full tribological response of rough, compliant interfaces under mixed and boundary lubrication regimes.

While two-scale frameworks represent a major step toward computationally feasible rough-contact modelling, they typically address the solid responses using classical half-space elasticity, which limits there applicability to soft, compliant materials. In such formulations, surface deformation is evaluated via Green's functions based on Flamant or Boussinesq solutions for one- and two-dimensional contacting interfaces, respectively \cite{Johnson1985}. Despite their computational efficiency, these models neglect finite-body effects and become unreliable when deformations are large relative to the contact geometry. To address these shortcomings, fully coupled finite element methods (FEMs) have been developed to incorporate non-linear material behaviour and large deformation \cite{stupkiewicz2009finite, stupkiewicz2016finite}. However, these models are computationally demanding and rarely include realistic surface roughness due to the high resolution required. Similarly, modified Green’s function formulations \cite{putignano2015mechanics, scaraggi2016effect} have been proposed to incorporate viscoelasticity within boundary element methods (BEMs), but they remain constrained by the half-space assumption. Consequently, a unified framework that can capture surface roughness, finite deformation, and non-linear elasticity within a tractable FSI formulation remains an open challenge in the modelling of lubricated compliant contacts.

Motivated by these limitations, the present study presents a robust, modular computational framework that efficiently integrates rough contact mechanics with finite surface deformation to accurately simulate the fluid-solid interactions driving compliant, rough contacting interfaces. Surface roughness is represented using both Persson's statistical theory \cite{persson2006contact} and a deterministic Conjugate Gradient with Fast Fourier Transform (CG-FFT) approach to efficiently address the multiscale complexity of real measured surface profiles. Furthermore, a novel Reduced Stiffness Method (RSM) is proposed to overcome the limitations of half-space assumptions, enabling accurate and computationally efficient treatment of finite-body elastic effects. The lubrication algorithm developed in \cite{persson2009transition} is extended to reproduce the complete evolution of friction with sliding speed (\emph{i.e.}, Stribeck curve profiles), capturing the transition from direct asperity contact to fully fluid-supported motion. The proposed multiscale FSI framework is validated against friction measurements on rough elastomer–glass interfaces, demonstrating its predictive capability across diverse operating regimes.

\begin{figure}[H]
        \centering
        \includegraphics[width=0.95\linewidth]{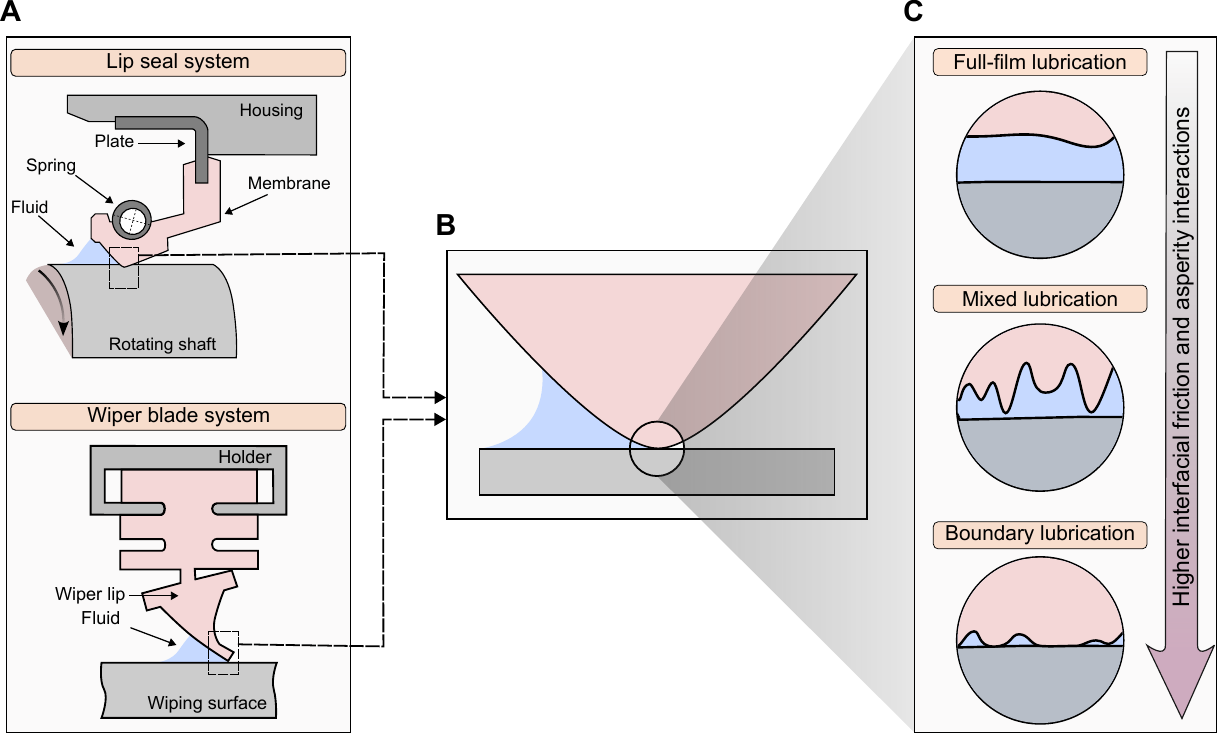}
        \caption{The modularity and robustness of the developed computational framework enable the modelling of diverse soft lubricated contacts and conditions. (\textbf{A})  Examples of soft lubricated interfaces, including wiper-blade and lip-seal configurations, exhibiting mixed lubrication between a compliant rubber lip and a rigid surface through coupled deformation and fluid entrainment; the wiper-blade system is investigated in this study as a representative case. (\textbf{B}) Simplified geometric model used for numerical analysis, where the complex wiper-blade profile is reduced to an elastomer specimen sliding against a rigid glass substrate. (\textbf{C}) Conceptual lubrication map showing the transition from full-film to boundary lubrication.}
        \label{fig:SoftContact_Configuration}
\end{figure}

This paper is structured as follows. \autoref{sec:Experiments} outlines the design of a tribological experimental setup inspired by automotive windscreen wiper systems, selected as a representative application where surface roughness, material compliance and lubrication interplay critically influence frictional performance. The complex wiper geometry is approximated by a simplified elastomer–glass configuration, as illustrated in \autoref{fig:SoftContact_Configuration}, which also serves as a generic model for other soft lubricated interfaces such as rotary lip and compliant shaft seals. \autoref{sec:Simulation} details the theoretical formulation and numerical implementation of the multiscale FSI framework, incorporating hydrodynamic flow, surface roughness and large elastic deformation. \autoref{sec:Results_Discussion} presents the simulation results and experimental validation, comparing Persson’s statistical theory and the CG–FFT method for rough-contact modelling. The framework predicts pressure, separation and friction over a range of sliding speeds and evaluates the performance of the proposed RSM approach relative to a conventional influence coefficient matrix approach. Finally, \autoref{sec:Conclusions} summarises the key findings and outlines future developments of the proposed FSI framework.

\section{Experimental Protocols}
\label{sec:Experiments}
\begin{figure}[H]
        \centering
        \includegraphics[width=0.9\linewidth]{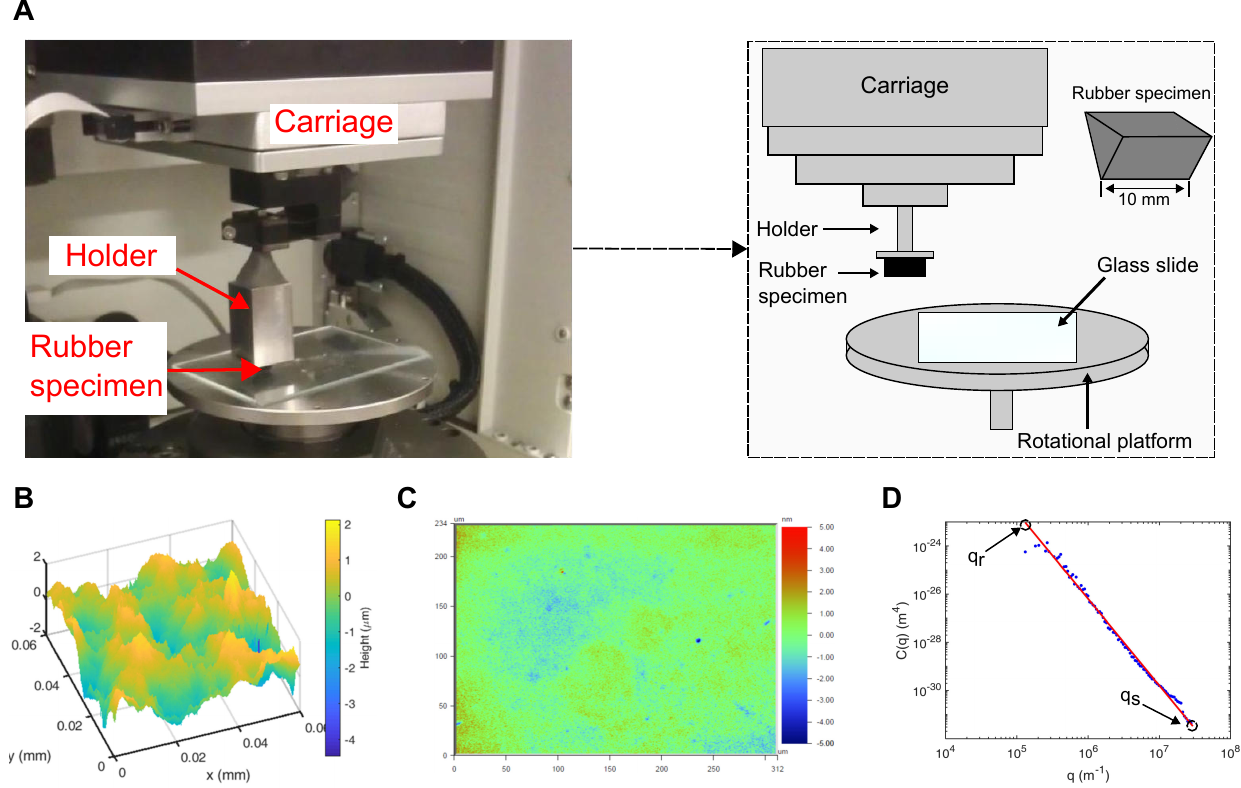}
        \caption{Experimental setup and surface characterisation for friction assessment. (\textbf{A}) The UMT2 tribometer with mounted elastomer specimen in contact with the glass slide. (\textbf{B}) Elastometer surface topography measured using Atomic Force Microscopy (AFM). (\textbf{C}) Glass substrate roughness map obtained via White Light Interferometry (WLI). (\textbf{D}) Power spectral density (PSD) of the elastomer surface, with the fitted roughness power spectrum $C({q})$ (in red) corresponding to a Hurst exponent of 0.8.}
        \label{fig:Experiment}
\end{figure}

A custom experimental platform inspired by automotive windscreen wiper systems was developed to corroborate the developed computational framework. The setup aims to reduce the geometric complexity of real wiper systems to a controlled configuration with a compliant elastomer specimen sliding against a rigid glass substrate (see \autoref{fig:Experiment}\textbf{A}). The elastomer specimens characterised by a Young’s modulus ($E$) of 3 MPa and a Poisson’s ratio ($\nu$) of 0.5, were fabricated with a triangular cross-section featuring a 0.2 mm tip radius. This geometry was selected to mitigate kinematic complexities such as large rotations while preserving the essential compliance and deformation behaviour representative of wiper-blade systems. Surface roughness measurements were obtained using Atomic Force Microscopy (AFM) as shown in \autoref{fig:Experiment}\textbf{B}, indicating a root-mean-square (RMS) roughness of 0.67 $\upmu\mathrm{m}$. Furthermore, glass microscope slides were used to simulate the windscreen surface due to its topographic uniformity and excellent cleanliness. The topography of the glass surface was measured using White Light Interferometry, as shown in \autoref{fig:Experiment}\textbf{C}, with an RMS roughness of approximately 0.5 nm. Due to the significant contrast in roughness scales, the glass surface was treated as rigid and nominally flat in both experiments and simulations.

Prior to testing, the glass substrates were cleaned using ultrasonic agitation in a toluene bath for 20 minutes, followed by rinsing in isopropanol and drying with a warm air stream. The elastomer samples were tested in their as-received condition, preserving their original hydrophobicity and surface finish. No additional surface treatments were applied to alter roughness or wettability. Tribological characterisation was carried out using a UMT2 tribometer, as illustrated in \autoref{fig:Experiment}\textbf{A}. The elastomer strip, 10 mm in width and mounted at a 17 mm radial offset from the centre, was held in a fixed horizontal orientation above the rotating glass disc. A controlled vertical load was applied to establish the desired nominal contact pressure, and the glass substrate was rotated at a series of prescribed angular velocities. Friction and normal forces were measured via calibrated transducers, enabling accurate evaluation of friction coefficients under various testing conditions.

Constructing a complete Stribeck curve profile, encompassing boundary, mixed and hydrodynamic lubrication regimes, entails spanning a wide range of Hersey numbers ($H = \eta UW^{-1}$), where $\eta$ is the dynamic viscosity of the lubricant, $U$ is the operating velocity and $W$ is the applied normal load. However, the tribometer’s maximum rotational speed constrains the range of attainable sliding velocities when using a single lubricant. To overcome this limitation, a set of test fluids with varying dynamic viscosities was employed, thereby enabling coverage of multiple Hersey number orders of magnitude. This approach follows similar methodologies previously adopted in studies pertaining to compliant contacts \cite{de2005frictional, myant2010influence}. The lubricants used in this work, along with their dynamic viscosities, are summarised in \autoref{tab:Lubricant_properites}. Viscosity measurements were conducted using a calibrated Stabinger viscometer (\textit{Anton Paar, UK}). New elastomer and glass specimens were used for each combination of lubricant and operating velocity in order to ensure reproducibility and eliminate cross-contamination. Following the completion of all test permutations, friction coefficients were plotted against the Hersey number to generate a full Stribeck curve, with the repeatability of these measurements and their comparison to numerical predictions discussed in \autoref{sec:Results_Discussion}.

\begin{table}[H]
        \centering
        \caption{Overview of the lubricants used in the tribological experiments, indicating their dynamic viscosities measured at 21\textdegree C and the corresponding velocity ranges evaluated during testing.}        
        \label{tab:Lubricants}
        \begin{tabular}{lccc}
            \toprule
            \textbf{Lubricant} & \textbf{Composition} & \textbf{Viscosity (mPa·s)} & \textbf{Velocity range (mm/s)} \\
            \midrule
            Water & Demineralised water   & 0.979  & 0.0178 \text{--} 178 \\
            GLY60 & 60 wt.\% Glycerol in water   & 10.8  & 0.178  \text{--} 53.4 \\
            GLY & Glycerol  & 306   & 1.8 \text{--} 178 \\
            \bottomrule
        \end{tabular}
        \label{tab:Lubricant_properites}
\end{table}

\section{Simulation Framework}
\label{sec:Simulation}
This section outlines the formulation and numerical implementation of the proposed FSI framework for lubricated, compliant and rough contacts. The model couples hydrodynamic lubrication, rough surface contact mechanics and large elastic deformation within a unified computational environment. At the macroscale, the global pressure distribution and deformation of the compliant body are resolved, while microscale roughness effects are incorporated locally through deterministic contact models. The formulation is developed in two dimensions, as the contact width is considerably smaller than the specimen length, which allows efficient computation without loss of accuracy. Nevertheless, the framework is readily extensible to three-dimensional configurations.

\subsection{Hydrodynamic Flow}
The interfacial fluid flow is governed by the incompressible, isoviscous and isothermal Reynolds equation expressed as follows \cite{Ardah2025}:
\begin{equation}
        \frac{\partial}{\partial x} \bigg( \frac{u^3}{6 \nu \eta} \frac{\partial p_f}{\partial x} \bigg) = \frac{\partial}{\partial x} (u),
        \label{eq:Reynolds}
\end{equation}
where $p_f$ is the hydrodynamic pressure, $u$ is the local film thickness, $\eta$ is the lubricant dynamic viscosity and $u$ is the tangential sliding velocity of the compliant body along the sliding direction (or $x$-axis). The film thickness $u$ is defined as the sum of the nominal gap and the local elastic deformation, adjusted by the contact geometry and surface topography.

The hydrodynamic pressure distribution is evaluated by discretising the Reynolds equation (\autoref{eq:Reynolds}) using the Finite Difference Method and iteratively obtaining the numerical solution via a Gauss–Seidel relaxation procedure. The computational domain, $[x_{\mathrm{in}}, x_{\mathrm{end}}]$, is chosen to be sufficiently long to achieve fully developed flow conditions and avoid inlet starvation, which is commonly observed in compliant lubricated contacts. Cavitation is represented using Gümbel’s model \cite{Gümbel1916}, or the half-Sommerfeld condition, by restricting the local pressure to values above the cavitation threshold $ p_{\mathrm{cav}}$.

\subsection{Rough Contact Mechanics}
\label{subsec:Rough_Contact_Mechanics}
The contact system considered in this study consists of a nominally flat rigid body possessing a one-dimensional rough surface profile, $z(\textbf{x})$, pressed against a flat elastic half-space characterised by Young’s modulus $E$ and Poisson’s ratio $\nu$. This configuration enables a thorough investigation of rough contact behaviour without the computational complexities of three-dimensional geometries. Surface roughness is statistically characterised by its Power Spectral Density (PSD) for a scale-invariant description of the surface morphology. As illustrated in \autoref{fig:Experiment}\textbf{D}, the PSD follows a power-law decay across spatial frequencies, a behaviour that is indicative of self-affine fractal surfaces that exhibit roughness over multiple length scales and are frequently encountered in both engineered and natural systems. For a self-affine fractal surface, the PSD takes the following form \cite{persson2006contact}:
\begin{align}
        C(\textbf{q}) = 
        \begin{cases}
           \dfrac{H}{\pi} \dfrac{{h_{rms}}^2}{{q_{r}}^{2}} \left( \dfrac{q}{q_r} \right)^{-2(H+1)} & \text{if} \quad \quad q_r < q < q_s
           \\[2ex]
           C_0 & \text{if} \quad \quad q_L < q < q_r,
        \end{cases}
\end{align}
where $\textbf{q}$ is the wavevector with magnitude $q = |\textbf{q}| = 2 \times {\pi}/{\lambda}$, $\lambda$ is the wavelength and $H$ is the Hurst exponent associated with the fractal dimension $D_f$ via $H = 3-D_f$. For the present analysis, the Hurst exponent is $H = 0.8$, with short- and long-wavelength cut-offs $q_s = 3 \times 10^7$ and $q_r = 1 \times 10^5$, respectively.

To address the contact mechanics of rough surfaces, two principal modelling strategies are commonly employed: stochastic theories and deterministic simulations. In the present study, both approaches are explored, in particular Persson’s statistical theory and a deterministic method based on the Conjugate Gradient (CG) algorithm. These methods are benchmarked against each other to assess their accuracy and applicability in predicting key interfacial quantities, such as the real contact area and the mean interfacial separation for accurately capturing the influence of surface roughness within the proposed computational framework.

\subsection{Persson’s Theory}
Consider an elastic body with a flat surface, characterised by Young’s modulus $E$ and Poisson’s ratio $\nu$, in contact with a rigid solid possessing a rough surface profile $z(x)$ under an applied pressure $\sigma_0$. Persson’s theory \cite{persson2001theory, persson2006contact} provides a statistical framework for describing the evolution of contact mechanics across multiple scales of magnification, denoted by $\zeta$. At low magnification ($\zeta = q/q_r$), where $q_r$ represents the reference wavevector corresponding to the longest resolvable wavelength, the surface appears smooth and the apparent contact is nearly complete at the macroscopic level. As the magnification increases, progressively shorter wavelength roughness components become resolved, leading to a reduction in the apparent contact area as contact localises over discrete asperities. Within this multiscale framework, Persson established a relationship between the magnification-dependent contact area ratio $A(\zeta)/A_0$ and the evolving interfacial stress probability distribution $P(\sigma, \zeta)$, where $\sigma$ denotes the local contact stress, expressed as follows:
\begin{equation}
    \frac{A(\zeta)}{A_0} = \int d\sigma \, \, P(\sigma, \zeta).
\end{equation}
When $\zeta = 1$, the solids appear to be in full contact, and the interfacial stress is uniformly equal to the applied pressure, resulting in a delta-function-like distribution given as follows:
\begin{equation}
    P(\sigma, 1) = \delta(\sigma - \sigma_0).
\end{equation}

As the magnification factor $\zeta$ increases, progressively finer surface features with shorter wavelengths become resolvable, revealing that contact occurs only between isolated asperities rather than across the entire nominal area. Therefore, the evolution of the interfacial stress distribution under increasing $\zeta$ follows a diffusion-like partial differential equation \cite{persson2006contact}, where the role of time is replaced by magnification $\zeta$ and the spatial coordinate is represented by the local stress $\sigma$, as expressed by the following:
\begin{equation}
    \frac{\partial P}{\partial \zeta} = f(\zeta) \frac{\partial^2 P}{\partial \sigma^2},
\end{equation}
where
\begin{equation}
    f(\zeta) = \frac{\pi}{4} \bigg( \frac{E}{1-\nu^2} \bigg) \, q_r \, q^3C(q),
\end{equation}
where $q_r = 2\pi/L$, with $L$ being the lateral size of the nominal contact area, and $q=\zeta q_r$. When suitable boundary conditions are imposed, Persson’s theoretical framework derives the following contact area ratio:
\begin{equation}
    \frac{A(\zeta)}{A_0} = \frac{1}{\sqrt{\pi}} \int_0^{\sqrt{G}} dx \, \, e^{-x^2/4} = \erf \bigg( {1}/{2 \sqrt{G}} \bigg),
\end{equation}
where
\begin{equation}
    G(\zeta) = \frac{\pi}{4} \bigg( \frac{E}{(1 - \nu^2) \sigma_0} \bigg) \int_{q_r}^{\zeta q_r} dq \, \, q^3C(q).
\end{equation}

Unlike the derivation of the contact area evolution, which is based on the assumption of full contact, the analysis of interfacial separation begins under conditions of small applied loads. Let $\bar{u} > 0$ denote the mean interfacial separation between the deformable elastic surface and the nominal (or average) plane of the rigid rough substrate. As the externally applied pressure $p$ increases, the average gap $\bar{u}$ decreases accordingly, reflecting the progressive engagement of surface asperities. The mechanical work expended by the external loading is then converted into stored elastic energy, which is predominantly concentrated in the vicinity of the microscale contact regions. This elastic energy, denoted by $U_{\text{el}}$, can be expressed as follows \cite{persson2007relation, yang2008contact}:
\begin{equation}
    U_{el}(\bar u) = \int_u^\infty d\bar{u} \, \, A_0 p(\bar{u}').
\end{equation}
In an approximate formulation, the elastic energy $U_{el}$ may be represented in its simplest form as follows:
\begin{equation}
    U_{el}(\bar u) \approx A_0 E_r \frac{\pi}{2} \gamma \int_{q_r}^{q_s} dq \, \, q^2 P(p) C(q).
\end{equation}
Given that both $P(p)$ and $C(q)$ are known, the mean interfacial separation between the two surfaces can be determined using:
\begin{equation}
    p(\bar{u}) = - \sqrt{\pi} \gamma \int_{q_r}^{q_s} dq \, \, q^2 C(q) s(q) e^{-\left[  {s(q) p}/{E_r} \right]^2} \frac{dp}{d\bar{u}},
    \label{eq:p_bar_u}
\end{equation}
or alternatively using:
\begin{equation}
    d\bar{u} = - \sqrt{\pi} \gamma \int_{q_r}^{q_s} dq \, \, q^2 C(q) s(q) e^{-\left[  {s(q) p}/{E_r} \right]^2} \frac{dp}{p},
    \label{eq:Eq20}
\end{equation}
where $s(q) = w(q)/E_r$. Therefore, an expression for $w(q)$ can be obtained as follows:
\begin{equation*}
    w(q) = \bigg( \pi \int_{q_r}^{q} dq' \, \, {q'}^3 C(q') \bigg)^{-1/2}.
\end{equation*}
It is important to note that a pre-factor $\gamma$ is introduced to account for the elastic energy stored within the contact regions and is suggested to be less than or equal to 1. At low squeezing pressures, Yang and Persson \cite{yang2008contact} employed molecular dynamics simulations to investigate the pressure–separation relationship $p(\bar{u})$ for self-affine fractal surfaces, reporting consistency with $\gamma \approx 0.4$.

For small applied loads, integrating \autoref{eq:Eq20} from $u = 0$ to $u$ yields an exact expression for the interfacial separation, given as follows \cite{yang2008contact}:
\begin{equation}
    u = \sqrt{\pi} \gamma \int_{q_r}^{q_s} dq \, \, q^2 C(q) s(q) \, \, \int_{p}^{\infty} dp' \, \, \frac{1}{p'} e^{-\left[ w(q) p'/E_r \right]^2}.
    \label{eq:Eq22}
\end{equation}
For purely self-affine fractal surfaces and very small pressures, \autoref{eq:Eq22} can be
further reduced to an approximated equation expressed as follows \cite{persson2007relation}:
\begin{equation}
    u = \beta_{1}^{-1} h_{rms} \log \left( \beta_2 \epsilon q_r h_{rms} E_{r} / p \right),
\end{equation}
where $\varepsilon$ is around 0.75, while $\beta_1$ and $\beta_2$ can be computed using the roughness power spectrum. For surfaces described by fractal dimension $D_f = 2.2$, $\beta_1$ and $\beta_2$ are 1.0 and 0.5 \cite{persson2007relation}, respectively.

\subsection{Deterministic Method}
While \autoref{eq:p_bar_u} offers a convenient means to estimate the contact area and interfacial separation of loaded systems, stochastic approaches are inherently reliant on statistical descriptors, such as the Hurst exponent ($H$) and reference wavevector ($q_r$), which may be difficult to accurately extract from power spectral density data and are limited to self-affine fractal surfaces. In contrast, deterministic approaches offer greater versatility, as they can be directly applied to arbitrary, experimentally measured surface topographies. In the following analysis, we adopt a deterministic approach based on the Boundary Element Method (BEM), under the assumption that the half-space approximation remains valid for the contact configuration under consideration.

Solving contact problems using BEM entails determining the elastic deformation of the solid under an applied pressure distribution. For two-dimensional rough surfaces, the classical elasticity theory provides that the deformation $w(x,y)$ at a point on the surface due to a distributed normal pressure $p$ can be evaluated as follows \cite{lorenz2010average}:
\begin{equation}
        w(x,y) = \frac{2}{\pi E_r} \bigintsss \!\!\!\!\! \bigintsss_{\Omega} \frac{{p}(\xi, \eta)}{\sqrt{\left( x - \xi \right)^{2} + \left( y - \eta \right)^{2}}} \,d\xi \, d\eta.
        \label{eq:Boussinesq}
\end{equation}
Alternatively, in its discretised form, \autoref{eq:Boussinesq} can be expressed as follows:
\begin{align}
    w(x_k, y_l) = \frac{2}{\pi E_r} \mathlarger{\mathlarger{\sum}}_{i=1}^{N_x} \, \mathlarger{\mathlarger{\sum}}_{j=1}^{N_y} D^{k,l}_{i,j} \, \, p(\xi_i, \eta_j),
    \label{eq:Boussinesq_Discrete}
\end{align}
where $D^{k,l}_{i,j}$ is the influence coefficient matrix which relates the deformation at nodal position $(k,l)$ with a load applied at nodal position $(i,j)$, while $N_x$ and $N_y$ denote the number of discretised elements along the x- and y-directions, respectively. Assuming a zero-order shape function, in which pressure is approximated as constant over each mesh element, the double integral admits an analytical solution that can be expressed as follows \cite{LoveTheSP}:
\begin{align}
     \begin{split}
        D^{k,l}_{i,j} &= \big| X_p \big| \, \mathrm{sinh^{-1}} \left( {\cfrac{Y_p}{X_p}} \right) + \big| Y_p \big| \, \mathrm{sinh^{-1}} \left( {\cfrac{X_p}{Y_p}} \right)
        \\
        &- \big| X_m \big| \, \mathrm{sinh^{-1}} \left( {\cfrac{Y_p}{X_m}} \right) - \big| Y_p \big| \, \mathrm{sinh^{-1}} \left( {\cfrac{X_m}{Y_p}} \right)
        \\
        &- \big| X_p \big| \, \mathrm{sinh^{-1}} \left( {\cfrac{Y_m}{X_p}} \right) - \big| Y_m \big| \, \mathrm{sinh^{-1}} \left( {\cfrac{X_p}{Y_m}} \right)
        \\
        &+ \big| X_m \big| \, \mathrm{sinh^{-1}} \left( {\cfrac{Y_m}{X_m}} \right) + \big| Y_m \big| \, \mathrm{sinh^{-1}} \left( {\cfrac{X_m}{Y_m}} \right),
        \label{eq:D_ijkl}
     \end{split}
\end{align}
where the relevant coefficients are defined in terms of the mesh spacings in the x- and y-directions, $\Delta x$ and $\Delta y$ respectively, as follows:
\begin{subequations} \label{eq:IC_coordinates}
     \begin{align}
        X_p &= x_k - x_i + \left( \Delta x \big/ 2 \right), \\
        X_m &= x_k - x_i - \left( \Delta x \big/ 2 \right), \\
        Y_p &= y_l - y_j + \left( \Delta y \big/ 2 \right), \\
        Y_m &= y_l - y_j - \left( \Delta y \big/ 2 \right).
     \end{align}
\end{subequations}        
To efficiently compute the surface deformation from a given pressure distribution, the present study employs the Discrete Convolution and Fast Fourier Transform (DC-FFT) method. This approach leverages the convolution theorem to significantly reduce the computational cost associated with evaluating the influence matrix sums in \autoref{eq:Boussinesq_Discrete}, particularly for large-scale two-dimensional problems. Readers interested in the underlying mathematical formulation and implementation of the DC-FFT technique are referred to \cite{Ardah2025, liu2000versatile, wang2020fft}.

A fundamental challenge in solving rough surface contact problems lies in accurately determining the real contact area. When a rough surface is compressed against a flat plane under an external load $W$, pressure develops in the contact regions, where the surfaces are in contact and the local separation vanishes, while non-contact regions are characterised by zero pressure and positive separation. Therefore, the total pressure integrated over the contact area must balance the applied load. Such physical scenario can be mathematically formulated through a system of equations and inequalities, as follows:
\begin{subequations}
    \label{eq:CG_Eqns}
    \begin{align}
        u_0 + u_{ij} + \sum_{(h,k) \in I_g} D^{k,l}_{i,j} \, p_{hk} &> 0 && \quad (i,j) \in I_c, \label{eq:CG_Eqn_1} \\[1ex]
        p_{ij} &> 0 && \quad (i,j) \in I_c, \label{eq:CG_Eqn_2} \\[1ex]
        u_0 + u_{ij} + \sum_{(h,k) \in I_g} D^{k,l}_{i,j} \, p_{hk} &> 0 && \quad (i,j) \notin I_c, \label{eq:CG_Eqn_3} \\[1ex]
        p_{ij} &= 0 && \quad (i,j) \notin I_c, \label{eq:CG_Eqn_4} \\[1ex]
        a_x a_y \sum_{(k,l) \in I_g} p_{ij} &= W, && \label{eq:CG_Eqn_5}
    \end{align}
\end{subequations}
where $u_0$ is the rigid-body approach, $u_{ij}$ is the surface roughness height, $I_c$ refers to the set of nodes in contact and $I_g$ encompasses all nodes within the computational domain.

According to the classical optimisation theory, \autoref{eq:CG_Eqn_1} to \autoref{eq:CG_Eqn_5} collectively define a quadratic complementarity problem subject to linear inequality constraints. Such formulations are particularly amenable to iterative solvers based on the Conjugate Gradient (CG) method, especially when suitably adapted to the structure of contact problems. A seminal implementation of this approach was introduced by Polonsky and Keer \cite{Polonsky1999Jul}, who applied the CG method to rough surface contact under non-periodic boundary conditions. Their algorithm proved effective in both load-controlled and displacement-controlled scenarios. In the present work, a load-controlled formulation is adopted, wherein global force equilibrium is explicitly enforced within each iteration cycle. This strategy not only accelerates convergence but also integrates naturally with fast convolution techniques such as the DC-FFT method. The resulting hybrid scheme allows for the efficient and robust solution of large-scale contact problems involving complex surface topographies. The adapted Conjugate Gradient (CG) method proceeds as follows. A full description of the algorithmic implementation of the CG-FFT approach including variable updates, inequality enforcement and load-balancing steps is provided in \ref{app:CGFFT}.

\begin{figure}[H]
        \centering
        \includegraphics[width=\linewidth]{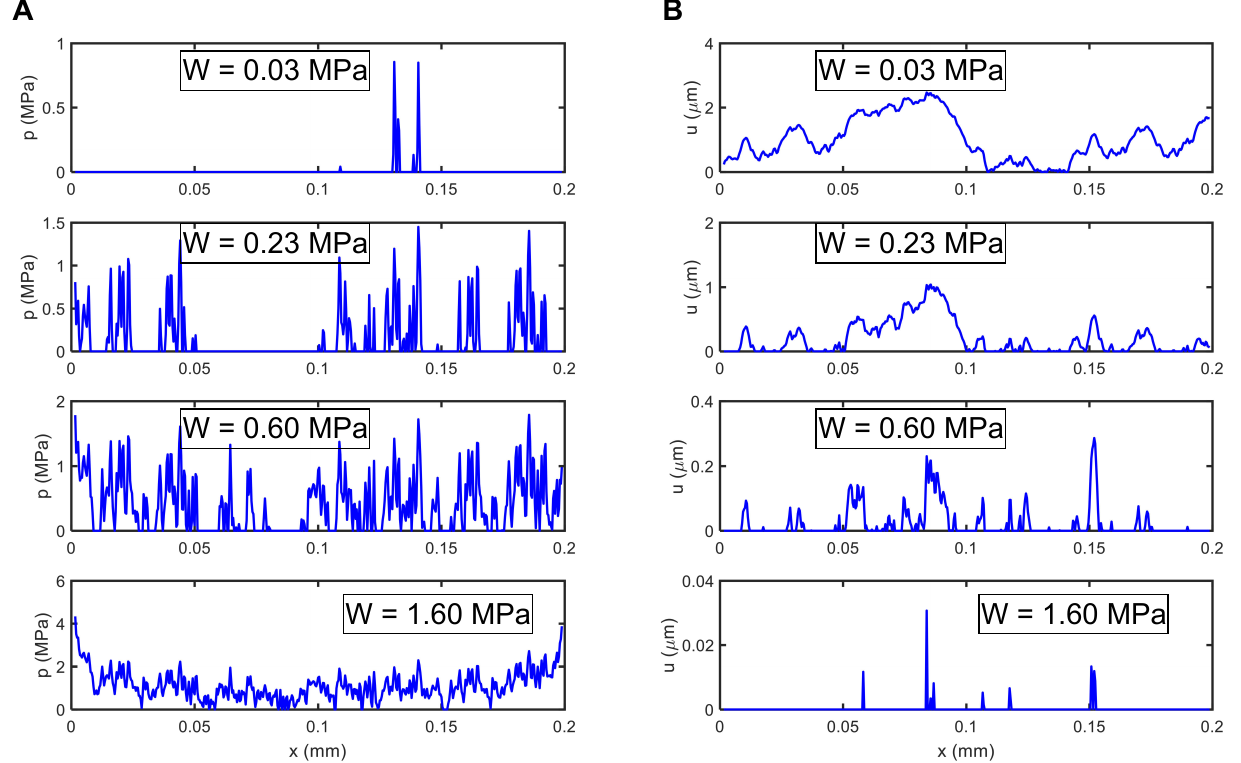}
        \caption{Evolution of the interfacial response of a rough elastic solid under increasing external load. Profiles along the central line show: (\textbf{A}) contact pressure and (\textbf{B}) interfacial separation. The material is characterised by a Young’s modulus of 3.3 MPa and a Poisson’s ratio of 0.499. Applied pressures are 0.03 MPa, 0.23 MPa, 0.6 MPa and 1.6 MPa.}
        \label{fig:Pressure_InterfacialGap}
\end{figure}

The CG-FFT algorithm was applied to synthetically generated rough surfaces, as shown in \autoref{fig:Experiment}\textbf{C}, and tested under varying normal loads. As with all FFT-based solvers, periodicity artefacts are inherently introduced due to the assumption of cyclic boundary conditions. To mitigate these artefacts and enable accurate treatment of non-periodic rough surfaces, the computational domain was extended to at least twice the nominal contact area using zero-padding and a wrap-around scheme. This ensures that edge effects are suppressed and interactions beyond the true contact region are sufficiently damped. Moreover, since the rough surfaces used in this study are significantly larger than the actual contact patches, residual periodicity errors are negligible.

\begin{figure}[H]
        \centering
        \includegraphics[width=\linewidth]{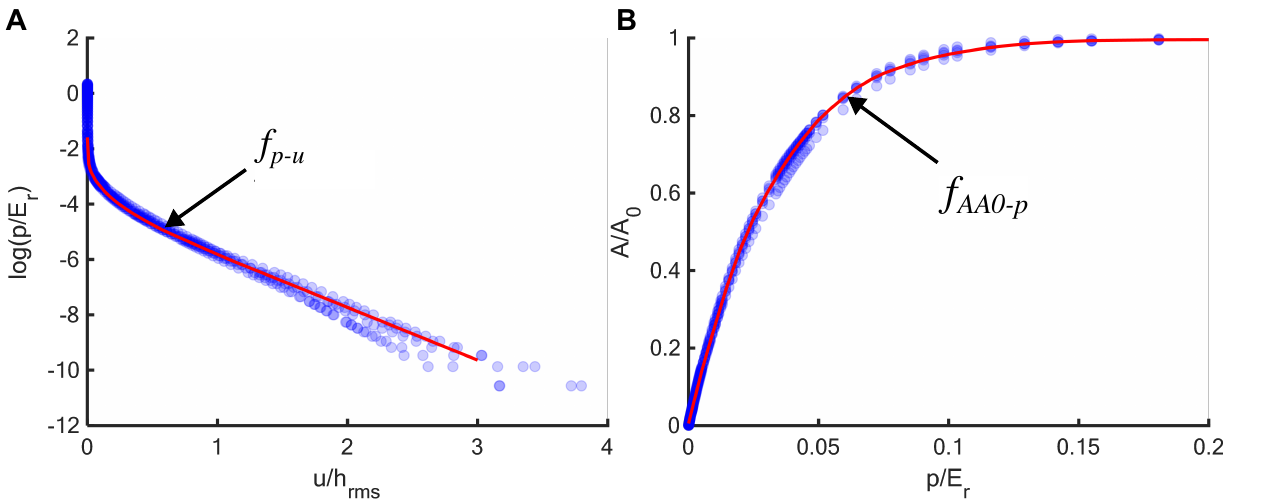}
        \caption{Functional relationships governing rough surface contact under varying load conditions. (\textbf{A}) Dimensionless load–separation curve extracted using the Conjugate Gradient with Fast Fourier Transform (CG-FFT) approach. (\textbf{B}) Contact area evolution with increasing normalised pressure, used in the construction of interpolation functions for pressure estimation and asperity-level stress evaluation.}
        \label{fig:Interpolation_Functions}
\end{figure}

To quantify the effect of external loading on contact behaviour, \autoref{fig:Pressure_InterfacialGap} presents the distribution of contact pressure and interfacial separation along the central line of a rough surface under varying applied loads using the CG-FFT approach. The results clearly show the increased contact pressure and reduced separation induced by higher external loading, consistent with theoretical expectations and prior literature. Furthermore, the average contact pressure $p$ and the mean interfacial gap $\bar{u}$ between the two surfaces can be computed based on those results. These quantities are then non-dimensionalised as $p/E_r$ and $\bar{u}/h_\text{rms}$, respectively, to establish the functional relationships between load and separation, as well as between the contact area ratio $A/A_0$ and pressure. These relationships are subsequently fitted with interpolation functions, denoted as $f_{p\text{-}u}$ and $f_{AA_0\text{-}p}$ as illustrated in \autoref{fig:Interpolation_Functions}\textbf{A} and \autoref{fig:Interpolation_Functions}\textbf{B}, respectively, which are later employed to estimate the contact pressure and evaluate shear stresses across asperity junctions, as detailed in \autoref{sec:Results_Discussion}.

\subsection{Surface Elastic Deformation}
\label{subsec:RSM}
The lubricant film thickness separating contacting interfaces in one-dimensional lubricated contacts can be expressed as follows:
\begin{equation}
   u(x) = u_0(x) + u_c + w(x),
\end{equation}
where $u_0$ is the profile prior to any deformation, $u_c$ is the central separation and $w$ is the elastic deformation. The elastic deformation can be determined by solving the convolution integral form given by the Flamant equation as follows \cite{Ardah2023Jan}:
\begin{align}
      {\delta}(x) = -\frac{2}{{\pi}E_{r}}\int_{x_{in}}^{x_{out}} p(\xi)\ln{(x-{s})}^2 \, \, ds,
      \label{eq:Flamant}
\end{align}
or in its respective discretised form given as:
\begin{align}
    w(x_k) = - \frac{2}{\pi E_r} \mathlarger{\mathlarger{\sum}}_{i=1}^{N_x}  I_{ki} \, p(\xi_i),
    \label{eq:Flamant_Discrete}
\end{align}
where $I_{ki}$ denotes the influence coefficient matrix for line contact configurations and represents the elastic displacement at a given node ($k$) resulting from a unit pressure applied at another node ($i$). While widely used, \autoref{eq:Flamant} and \autoref{eq:Flamant_Discrete} are based on the assumptions of linear elasticity and the half-space model, and are only valid when deformations are small relative to the overall geometry \cite{Ardah2025}. Under such conditions, deflections depend solely on the applied external load and not on the surrounding boundary conditions. However, these assumptions often break down in the context of soft materials, where finite deformation effects are significant. To address this limitation, we introduce the Reduced Stiffness Method (RSM), a more accurate and computationally efficient alternative which accounts for large deformations and material compliance more effectively.

In structural analysis, a standard approach for computing the deformation of a finite body involves solving the equations of motion derived from the finite element (FE) discretisation of the underlying continuum model, given as \cite{Ardah2025}:
\begin{align}
    \bigl[ \textbf{M} \bigr] \bigl\{ \boldsymbol{\ddot{{u}}} \bigr\} + \bigl[ \textbf{B} \bigr] \bigl\{ \boldsymbol{\dot{{u}}} \bigr\} + \bigl[ \textbf{K} \bigr] \bigl\{ \boldsymbol{{{u}}} \bigr\} = \bigl\{ \boldsymbol{{{f}}} \bigr\},
    \label{eq:FEM}
\end{align}
where $\bigl[ \textbf{M} \bigr]$, $ \bigl[\textbf{B} \bigr]$ and $\bigl[ \textbf{K} \bigr]$ are the mass, damping and stiffness matrices of the full FE model corresponding to the structural assembly, respectively. The vectors $\bigl\{ \boldsymbol{\ddot{{u}}} \bigr\}$, $\bigl\{ \boldsymbol{\dot{{u}}} \bigr\}$ and $\bigl\{ \boldsymbol{{{u}}} \bigr\}$ represent the nodal accelerations, velocities and displacements, respectively, while $\bigl\{ \boldsymbol{{{f}}} \bigr\}$ represents the applied nodal forces. These vector quantities encompass all the three-dimensional degrees of freedom for each node of the entire FE model.

As the primary interest lies in the deformation profiles at the contact interface, it is advantageous to apply substructuring techniques that condense the full FE model to an equivalent system that \emph{retains} only the degrees of freedom associated with the nodes on the contact surface. The \textit{reduced} form of \autoref{eq:FEM} can be expressed as:
\begin{align}
    \bigl[ \textbf{M}_\textbf{r} \bigr] \bigl\{ \boldsymbol{\ddot{{u}}_\textbf{r}} \bigr\} + \bigl[ \textbf{B}_\textbf{r} \bigr] \bigl\{ \boldsymbol{\dot{{u}}_\textbf{r}} \bigr\} + \bigl[ \textbf{K}_\textbf{r} \bigr] \bigl\{ \boldsymbol{{{u}}_\textbf{r}} \bigr\} = \bigl\{ \boldsymbol{{{f}}_\textbf{r}} \bigr\},
    \label{eq:FEM_reduced}
\end{align}
where the subscript $r$ denotes quantities corresponding to the reduced system. It is important to notice that, unlike the sparse matrices of the full FE model, the matrices of the reduced system are fully populated due to the inherent coupling of all degrees of freedom in substructure systems (super-elements). Additionally, any boundary conditions defined for the full FE model are automatically incorporated and enforced in the condensed model. Assuming a quasi-static condition and neglecting both inertial and damping effects, \autoref{eq:FEM_reduced} further simplifies to:
\begin{align}
    \bigl[ \textbf{K}_\textbf{r} \bigr] \bigl\{ \boldsymbol{{{u}}_\textbf{r}} \bigr\} = \bigl\{ \boldsymbol{{{f}}_\textbf{r}} \bigr\}.
    \label{eq:FEM_reduced_V2}
\end{align}

In this study, the reduced stiffness matrix $\left[ \mathbf{K}_\mathrm{r} \right]$ is generated through model reduction (or model condensation) using the substructure generation functionality available in Abaqus. This approach condenses a large, complex finite element system into a smaller, computationally efficient representation by retaining only a subset of degrees of freedom at key interfacial nodes in the computational domain. The process offers several advantages, such as a significant reduction in matrix size, accelerated matrix operations and automatic incorporation of boundary conditions and material properties defined in the full model.

\begin{figure}[H]
        \centering
        \includegraphics[width=\linewidth]{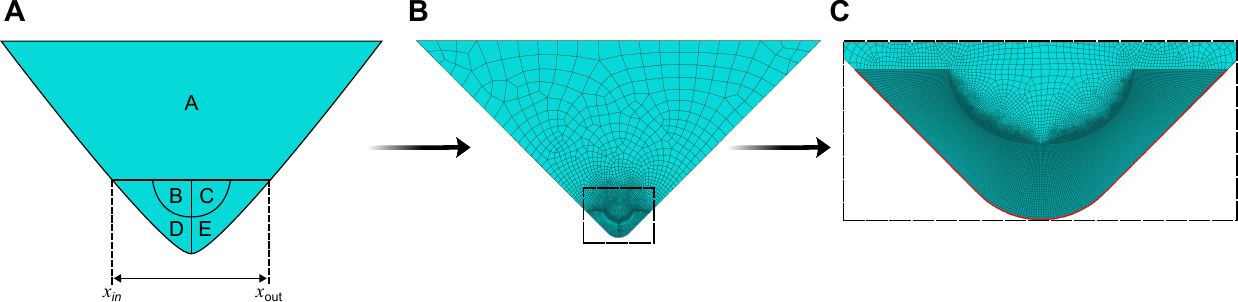}
        \caption{Schematic of the model reduction process involved in the Reduced Stiffness Matrix (RSM) method. (\textbf{A}) The full continuum geometry. (\textbf{B}) The corresponding finite element mesh with adaptive refinement. (\textbf{C}) The resulting reduced model containing retained nodes (in red) that define external degrees of freedom during substructure generation.}
        \label{fig:Reduced_Structure_Matrix}
\end{figure}

A schematic overview of the workflow is provided in \autoref{fig:Reduced_Structure_Matrix}, and the procedure is summarised below. A two-dimensional model of the elastic solid is first created and partitioned into several regions (see \autoref{fig:Reduced_Structure_Matrix}\textbf{A}), enabling adaptive mesh refinement across the domain (see \autoref{fig:Reduced_Structure_Matrix}\textbf{B}). This partitioning allows for fine, structured meshing near the contact interface, specifically in subdomains 'D' and 'E' where high spatial resolution is required to capture the pressure-induced deformation accurately. Uniform node spacing along the contact edge can be enforced to ensure compatibility with the hydrodynamic fluid solver within the computational length $[x_{\text{in}}, x_{\text{end}}]$ that encompasses the fluid domain. Elsewhere, the mesh is gradually coarsened using quad-dominated elements to balance accuracy with computational efficiency as showcased in \autoref{fig:Reduced_Structure_Matrix}\textbf{C}.

Once the mesh is defined, material properties and boundary conditions can be assigned to reflect the physical constraints relevant to substructure behaviour, with the top edge of the model fixed to prevent rigid-body motion. Substructure generation is performed using a linear perturbation analysis in Abaqus, producing a reduced stiffness matrix $\left[ \mathbf{K}\mathrm{r} \right]$ that captures the system’s linear response about a defined base state, either deformed or initial, depending on the analysis setup, with the initial response denoted as $\left[ \mathbf{K}\mathrm{r}^0 \right]$. Retained degrees of freedom, typically the first two translational components at nodes along the contact interface, are specified to ensure accurate deformation representation while eliminating internal degrees of freedom. The resulting matrix enables efficient and accurate deformation calculations within the broader FSI framework. This Reduced Stiffness Method (RSM), grounded in the Guyan condensation approach \cite{craig2006fundamentals}, offers significant computational advantages while preserving the essential physics of the full FE model.

It is important to note that in a linear perturbation analysis, the model’s response is governed by the linear elastic stiffness matrix evaluated at the base state, which is the configuration achieved at the end of the last general analysis step preceding the perturbation step. As a result, all external constraints present at the base state such as the boundary conditions, interaction definitions and loading states are inherently incorporated into the reduced stiffness matrix. If the linear perturbation step is defined as the first step of the analysis, the model is assumed to remain in its undeformed configuration, and only the initial boundary conditions are considered. In this case, the extracted stiffness matrix reflects the structural response under zero external load and is referred to as the initial stiffness matrix, denoted as $\bigl[ \textbf{K}_\textbf{r}^\textbf{0} \bigr]$.

\subsection{Computational Algorithm}
The FSI framework developed in this study comprises two computational modules that capture the strong coupling between fluid pressure and elastic deformation governed by the Reynolds equation: the Boundary–Mixed Lubrication (BML) module and the Mixed–Elastohydrodynamic Lubrication (MEHL) module. While the original algorithm introduced in \cite{persson2009transition} was formulated for point contacts to simulate transitions between boundary and full-film lubrication regimes, the present work extends this foundation by integrating a rough contact formulation and introducing the Reduced Stiffness Method to account for finite-body deformation effects in compliant solids.

The overall workflow for conducting a full FSI analysis begins with the procssing of experimentally measured rough surfaces to solve the normal contact problem via the deterministic CG–FFT method. The resulting contact characteristics, specifically the fitted pressure–gap relationships $f_{p\text{-}u}$, are then used to define normal contact within the FSI framework. Subsequently, FE analyses are carried out to simulate the sliding motion of the compliant system against a rigid substrate (\emph{i.e}, elastomer), from which the initial reduced stiffness matrix $\left[ \mathbf{K}_\mathrm{r}^0 \right]$ and the equivalent compliance matrix $\left[ \mathbf{EK}^{0} \right]_{ij}$ are extracted using the RSM approach, as detailed in \ref{app:MEHL}. These key inputs, namely, the fitted contact functions $f_{p\text{-}u}$, the reduced stiffness matrix $\left[ \mathbf{K}_\mathrm{r}^0 \right]$, and the undeformed shape of the elastomer $u_0$, are used to initialise the main FSI solver. Within this framework, both the BML and MEHL modules are updated to incorporate these quantities.

\begin{figure}[H]
        \centering
        \includegraphics[width=\linewidth]{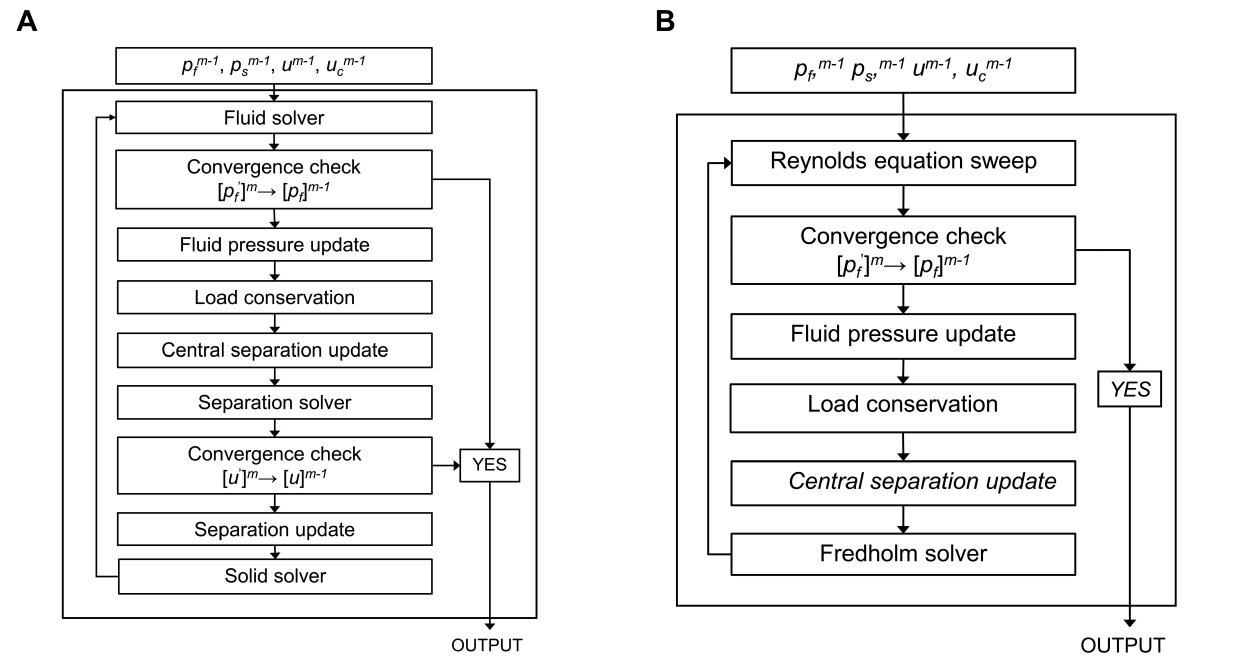}
        \caption{Overview of the computational workflow used in the multiscale FSI framework. The flowchart illustrates the iterative solution procedure for (\textbf{A}) the Boundary–Mixed Lubrication (BML) module and (\textbf{B}) the Mixed–Elastohydrodynamic Lubrication (MEHL) module.
}
        \label{fig:BML_MEHL}
\end{figure}

\subsubsection{Boundary-Mixed Lubrication}
In the Boundary–Mixed Lubrication (BML) module, the fluid pressure $p_f$, the solid contact pressure $p_s$ and the interfacial separation $u$ are computed iteratively and updated until convergence is achieved, as illustrated in \autoref{fig:BML_MEHL}\textbf{A}. This iterative coupling reflects the strong interdependence between hydrodynamic and solid-contact forces in partially lubricated interfaces. The solution process begins with an initial guess for the pressure distribution, which may either follow a Hertzian-like profile or be taken from the converged solution at a another sliding velocity condition. Using this initial guess, the hydrodynamic pressure $p_f$ is computed by solving the Reynolds equation (\autoref{eq:Reynolds}) within the fluid solver. After each iteration, the pressure field is updated using an under-relaxation strategy to enhance numerical stability, and convergence is systematically checked before advancing to the next stage of the algorithm as follows:
\begin{equation}
   p_f^m = p_f^{m-1} + \alpha_f \bigg( p_f'^m - p_f ^{m-1} \bigg),
\end{equation}
where $\alpha_f$ is the relaxation factor and the superscript $m$ indicates the iteration time step.

Load conservation is then enforced by rescaling the solid pressure, yielding a temporary renormalised value $p_s^{\text{temp}} = k_1 p_s$, where the scaling factor $k_1$ is defined as follows:
\begin{equation}
   k_1 = \frac{W - \mathlarger\int p_f^m(x') \, \, dx'}{\mathlarger\int p_s^{m-1}(x') \, \, dx'}.
\end{equation}

In the boundary lubrication regime, where solid contact pressure predominates within the contact zone, the minimum interfacial separation is assumed to occur at the centre of the contact, x = 0. This central interfacial separation can be directly inferred from the pertaining solid pressure using Persson’s contact theory as follows \cite{persson2007relation}:
\begin{equation}
    u_c = h_{rms} \ln \bigg( \frac{0.375 \times E_r q_r h_{rms}}{{p_s^\text{temp}}_{(\text{at }x=0)}} \bigg).
\end{equation}
Based on the rescaled pressure $p_{\text{total}} = p_s^{\text{temp}} + p_f$ and the central separation $u_c$, the separation between the contacting surfaces $u$ is evaluated from the elastic integral equation, before another convergence check and update for separation using relaxation factor $\alpha_u$. Accordingly the solid pressure is calculated using Persson’s analytical expression under-relaxed using $\alpha_s$ as follows:
\begin{equation}
    p_s \approx 0.375 \times E_r q_r h_{rms} e^{(u(x) / h_{rms})}.
\end{equation}

\begin{figure}[H]
        \centering
        \includegraphics[width=\linewidth]{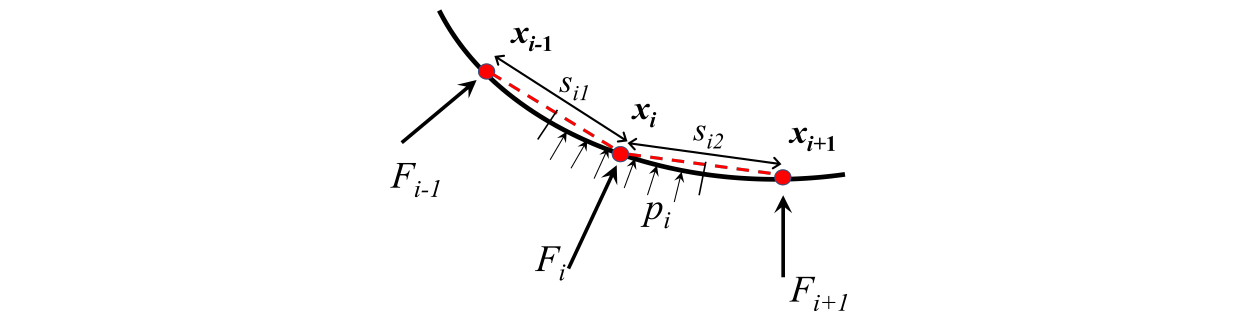}
        \caption{Schematic representation of the nodal force distribution and applied pressure field in the reduced stiffness model.}
        \label{fig:Force_RSM}
\end{figure}

To compute the deformation using the RSM approach, the rescaled total pressure field $p_{\text{total}}$ must first be converted into nodal force vectors. Mathematically, forces on each mesh node can be acquired by integrating pressure acting on the element. In the context of the two-dimensional model, if the pressure is assumed to be uniformly distributed around the node and the linear interpolation function is adopted, the line load acting at node $i$ can be expressed as:
\begin{equation}
   \boldsymbol{\mathrm{F}}_i = p_i \times \big(0.5 \Delta s_{i1} + 0.5 \Delta s_{i2} \big) \cdot \boldsymbol{\mathrm{n}}_e,
\end{equation}
where $\Delta s_{i1}$ and $\Delta s_{i2}$ represent the distances between $x_{i-1}$ and $x_i$, and between $x_i$ and $x_{i+1}$, respectively, as illustrated in \autoref{fig:Force_RSM}.

Subsequently, the nodal displacements are computed using \autoref{eq:FEM_reduced_V2}, where the initial stiffness matrix $\bigl[ \textbf{K}_\textbf{r}^{0} \bigr]$, extracted from the finite element model, is combined with the force vector $\boldsymbol{\mathrm{F}}_i$, obtained by integrating the total pressure field. It is important to note that the resulting displacement vector includes components in both directions; however, only the vertical displacements are retained for the deformation analysis. Thereafter, the solid pressure $p_s$ is updated based on the newly computed separation $u$ using the fitted interpolation function $f_{p-u}$, and subsequently relaxed using a factor $\alpha_s$. The iterative process continues until convergence is achieved, defined as the condition in which the residuals of both the fluid pressure $p_f$ and the separation $u$ fall below a prescribed threshold.

\subsubsection{Mixed-Hydrodynamic Lubrication}
\label{subsec:MEHL}
As the system transitions into the mixed lubrication regime, increased entrainment of lubricant leads to a rise in fluid pressure, which becomes comparable in magnitude to the solid contact pressure. This evolution introduces strong coupling between fluid pressure and interfacial separation, rendering the previously described BML algorithm inadequate and often prone to convergence difficulties. To overcome these limitations, the Mixed–Elastohydrodynamic Lubrication (MEHL) module is employed to capture the progression from partial to full-film lubrication conditions with improved numerical stability.

The MEHL framework integrates three key computational components, namely: (1) the Reynolds Equation Sweep (RES) for fluid pressure updates, (2) the Central Separation Updater (CSU) for dynamic film thickness adjustment, and (3) the Fredholm Solver (FS) for resolving elastic deformation and contact pressure (see \autoref{fig:BML_MEHL}\textbf{B}). Unlike the BML solver which fully resolves the Reynolds equation for a prescribed separation profile, the RES solver performs a single Gauss–Seidel sweep over the original Newton–Raphson scheme to avoid singularities in the Jacobian. After each pressure update, the CSU solver adjusts the central separation according to a load-balancing criterion, ensuring convergence toward the prescribed external load. The FS solver then iteratively resolves a Fredholm-type integral equation to obtain the interfacial separation profile, where the solid contact pressure is treated as a non-linear function of the local separation. A detailed step-by-step implementation of the MEHL algorithm, including all governing equations, update procedures and convergence criteria, is provided in \ref{app:MEHL}.

\subsection{Evaluation of Friction Response}
The frictional behaviour across different lubrication regimes can be predicted from the pressure distribution and interfacial separation obtained via the developed FSI computational model. The total interfacial shear stress, $\sigma_{\text{total}}$, comprises contributions from fluid shear ($\sigma_{{f}}$) and solid–solid contact ($\sigma_{{s}}$), and is expressed as follows:
\begin{equation}
    \sigma_{\text{total}} = \sigma_{\text{f}} + \sigma_{\text{s}}.
\end{equation}
The fluid shear stress component, $\sigma_{\text{f}}$, arises from the lubricant’s resistance to shearing. Assuming an iso-viscous fluid, $\sigma_{\text{f}}$ depends on the local velocity gradient and pressure distribution based given as follows:
\begin{equation}
    \sigma_{\text{f}} = \eta \frac{\partial v}{\partial z} = \eta \frac{v}{u(x)} + \frac{u(x)}{2} \frac{\partial p_f}{\partial x}.
\end{equation}
On the other hand, the solid–solid shear contribution, $\sigma_{\text{s}}$, is defined in terms of the interfacial shear strength $\sigma_{\text{1}}$ and the ratio of real to nominal contact area, $A/A_0$, as numerically described here \cite{persson2009transition}:
\begin{equation}
    \sigma_{\text{s}} = \sigma_{\text{1}} \frac{A}{A_0},
\end{equation}
where $\sigma_{\text{1}}$ represents the friction from van De Waals interactions and is approximated as a constant fraction of the elastic modulus, expressed as $\sigma_{\text{1}} = aE$, where $a$ is a dimensionless coefficient and $E$ is the Young’s modulus of the soft body. While $a$ is treated as a constant in the original formulation \cite{persson2009transition}, experimental studies have indicated that it may vary with sliding velocity \cite{vorvolakos2003effects}. Moreover, recent studies have highlighted that $\sigma_{\text{1}}$ is not solely governed by mechanical deformation but is strongly influenced by the chemical and physical nature of the interface, such as adhesive bonding, molecular layering and tribochecimal reactions that evolve under operational stresses and temperatures \cite{Ardah2025}. These interactions can modify the interfacial energy, alter the adhesion forces and even induce phase transitions within confine films, particularity in soft or polymeric materials. Therefore, both the magnitude and evolution of $\sigma_{\text{s}}$ is contingent on the interplay between contact mechanics, interfacial chemistry and operating conditions. The real contact area ratio, $A/A_0$, is computed from the solid pressure using the interpolation function $f_{{AA}_0{-}p}$, providing a direct link between pressure distribution and interfacial friction as shown in \autoref{fig:Interpolation_Functions}\textbf{B}.

Therefore, the total friction force can be computed by integrating the total shear stress as follows:
\begin{equation}
    F_{\text{total}} = \int \big( \sigma_{\text{f}} + \sigma_{\text{s}} \big) \, \, dx,
\end{equation}
while the coefficient of friction is evaluated according to the following relationship:
\begin{equation}
    \mu = \frac{F_{\text{total}}}{F_{\text{n}}} = \frac{\mathlarger{\mathlarger{\int}} \sigma_{\text{total}} \, \, dx}{\mathlarger{\mathlarger{\int}} p_{\text{total}} \, \, dx}.
\end{equation}

\section{Results and Discussion}
\label{sec:Results_Discussion}
This section is organised into two main parts, each designed to demonstrate the performance, accuracy, and practical relevance of the proposed framework. First, the fundamental behaviour of rough surface contacts is examined by comparing Persson’s statistical theory with the deterministic CG–FFT method. Second, numerical predictions are validated against experimental measurements, focusing on pressure distribution, film thickness, and frictional response across distinct lubrication regimes. The influence of different elastic deformation strategies is investigated by benchmarking the Reduced Stiffness Method (RSM) against the conventional influence matrix approach, with a detailed assessment of the RSM formulation and its comparison with the influence coefficient method provided in \ref{app:FFT_IC}.

\subsection{Numerical Validation of Rough Contact Behaviour}
\begin{figure}[H]
        \centering
        \includegraphics[width=\linewidth]{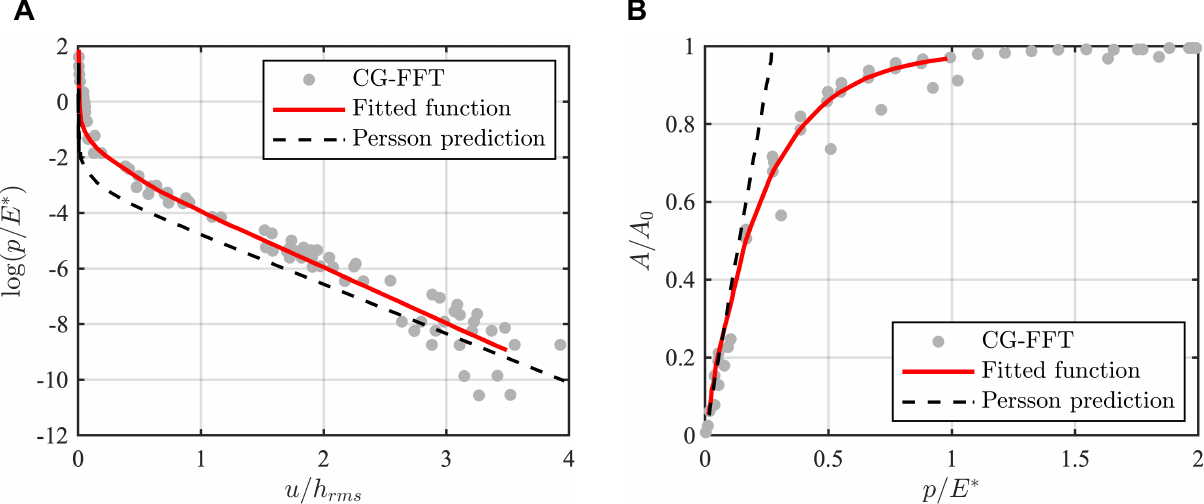}
        \caption{Comparison between Persson’s statistical theory and the deterministic CG–FFT method for modelling rough contact behaviour. (\textbf{A}) Load–separation relationship curves. (\textbf{B}) Contact area evolution curves.}
        \label{fig:Persson_CGFFT}
\end{figure}

A comparative analysis is conducted between Persson’s statistical theory and the deterministic Conjugate Gradient with Fast Fourier Transform (CG–FFT) method to evaluate their effectiveness in modelling rough contact behaviour. \autoref{fig:Persson_CGFFT} shows the variation of interfacial separation and real contact area ratio with applied load for the rough surface configuration detailed in \autoref{subsec:Rough_Contact_Mechanics}.

In order to achieve statistical robustness, three independent rough surfaces sharing an identical power spectral density, shown in \autoref{fig:Experiment}\textbf{D}), were analysed using the CG-FFT framework. The resulting  data (grey points) demonstrate excellent consistency with theoretical trends predicted by Persson’s model. Notably, while slight deviations emerge at very low loads, the CG–FFT results accurately capture the exponential decay of mean pressure with increasing interfacial separation within the intermediate loading range. More importantly, the method successfully reproduces the strongly non-linear mechanical response at small separations, a regime dominated by asperity interactions and elastic confinement, crucial for resolving the pressure–separation coupling that governs mixed and boundary lubrication dynamics. At large separations ($u > 3 \times h_{\mathrm{rms}}$), statistical fluctuations dominate, and data from this regime are therefore excluded from subsequent calibration. The remaining dataset is employed to derive smooth interpolation functions, $f_{p\text{–}u}$ and $f_{AA_0\text{–}p}$ (shown by the red curves in \autoref{fig:Persson_CGFFT}), which quantitatively link pressure, separation and real contact area. These functions provide a direct, data-driven bridge between the microscale contact mechanics and the macroscale hydrodynamic model within the FSI framework, enabling efficient yet accurate coupling between the solid deformation and fluid pressure fields.

Furthermore, predictions obtained using Persson’s statistical theory are shown as black dashed lines in \autoref{fig:Persson_CGFFT}. To achieve optimal agreement with the CG–FFT results, the prefactor $\gamma$ in \autoref{eq:p_bar_u}, originally reported as 0.48 in \cite{persson2007relation}, was empirically adjusted to unity. Such calibration, often required to account for surface-specific spectral characteristics, is consistent with prior findings \cite{papangelo2017load, prodanov2014contact}. Despite this tuning, Persson’s theory systematically underestimates the mean interfacial separation at low loads, particularly within the onset of contact where asperity-scale deformation dominates. This discrepancy highlights a key limitation of statistical approaches, as they do not account for local mechanical correlations or finite-size effects that influence real interfaces. These effects are inherently captured by deterministic solvers such as CG-FFT, and become more pronounced when utilising the RSM which explicitly incorporates finite geometry deformation.

Overall, the comparison underscores that while Persson’s theory provides valuable analytical scaling and physical intuition, the CG–FFT approach offers a superior representation of near-contact phenomena essential for predictive fluid–structure interaction modelling. By accurately capturing both the elastic nonlinearity and topographic heterogeneity of rough surfaces, the CG–FFT method establishes a reliable foundation for constructing reduced-order models of mixed lubrication within the developed multiscale framework. Beyond the present study, these insights emphasise the importance of coupling deterministic rough contact solvers with continuum-scale hydrodynamic formulations to achieve truly predictive simulations of lubricated soft interfaces. The demonstrated ability to reconcile microscale asperity mechanics with macroscopic frictional response provides a transferable modelling pathway for a wide class of soft interfacial system where surface roughness plays a critical role in regulating interfacial transport and dissipation.

\subsection{Experimental Validation}
\label{subsec:Exp_Validation}
Numerical simulations were performed under conditions replicating the experimental configuration described in \autoref{sec:Experiments} to assess the accuracy and robustness of the proposed multiscale FSI framework. The load–separation relationship, $f_{p\text{-}u}$, derived from the deterministic CG–FFT approach, and the reduced stiffness matrix obtained via the finite element analysis (see \autoref{subsec:RSM}), were integrated into the coupled solver. A normal load of 17 N/m was applied, and the lubricant was modelled as iso-viscous with a dynamic viscosity of 1 mPas. Sliding velocities spanned from $\mathrm{10^{-5}}$ m/s to 5 m/s, covering the entire operational range deployed in the experiments. The computational domain was discretised into 513 nodes, a resolution determined from mesh convergence tests to ensure accurate representation of pressure gradients and deformation fields while maintaining numerical efficiency.

\begin{figure}[H]
        \centering
        \includegraphics[width=\linewidth]{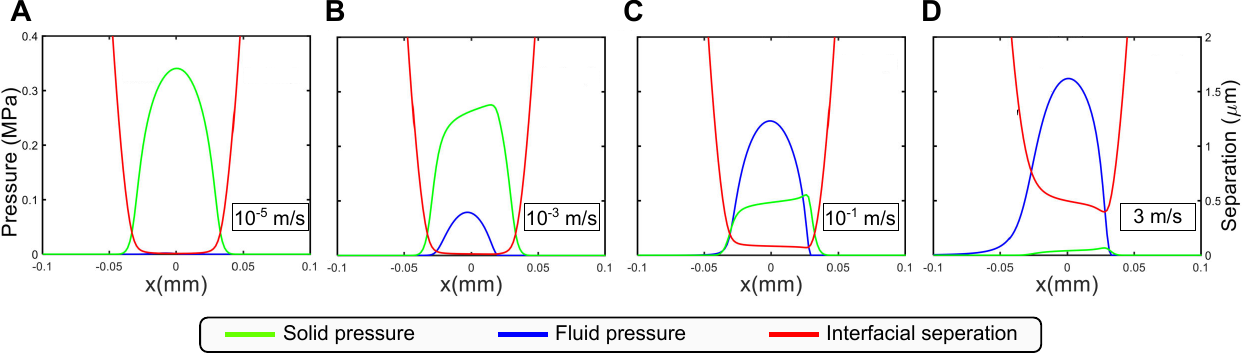}
        \caption{Predicted evolution of fluid pressure, solid pressure and interfacial separation with increasing sliding velocity. The transition from solid-dominated to fluid-supported contact as hydrodynamic pressure builds up is showcased with different sliding speeds (in m/s): (\textbf{A}) $10^{-5}$, (\textbf{B}) $10^{-3}$, (\textbf{C}) $10^{-1}$, and (\textbf{D}) 3.}
        \label{fig:PressureField}
\end{figure}

The pressure distribution and interfacial separation profiles at four representative sliding speeds are shown in \autoref{fig:PressureField}. As the sliding velocity increases, enhanced lubricant entrainment leads to a pronounced build-up of hydrodynamic pressure within the contact zone, resulting in a progressive transition from predominantly solid-supported contact at low speeds to a fully fluid-supported regime at higher velocities. At 3 m/s, the majority of the load is carried by the fluid film, with the contribution from solid contact becoming negligible, hence marking the onset of full-film lubrication regime. The predicted film thicknesses range from approximately 0.1 $\upmu\mathrm{m}$ at low velocities to 0.5 $\upmu\mathrm{m}$ at higher speeds, closely matching interferometric measurements reported for similar wiper-blade systems \cite{dobre2018tribological}. These results confirm that the model accurately captures the balance between surface roughness effects, material compliance and hydrodynamic pressure generation governing mixed lubrication dynamics.

\begin{figure}[H]
        \centering
        \includegraphics[width=\linewidth]{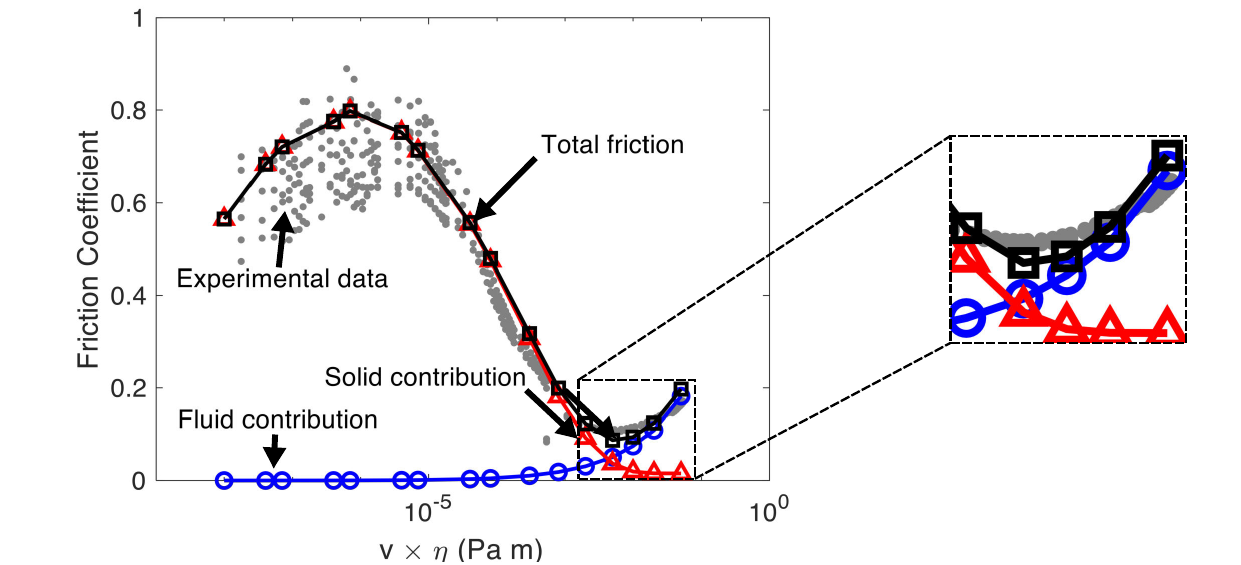}
        \caption{Validation of the developed FSI framework against experimental data for a compliant, rough lubricated contact. Comparison between experimentally measured (grey scatter points) and numerically predicted Stribeck curves obtained under an applied load of 17 N/m. The model accurately captures the transition from boundary to mixed and full-film lubrication regimes, highlighting fluid (blue), solid (red) and total (black) friction contributions.}
        \label{fig:StribeckCurve}
\end{figure}

Using the computed pressure and film separation profiles, the friction coefficient was evaluated and compared against experimental data in \autoref{fig:StribeckCurve}. The numerical predictions reproduce the classical Stribeck behaviour \cite{Ardah2025}, capturing the distinct transition from boundary lubrication at low speeds, through mixed lubrication, to fully fluid-supported motion at high speeds. The fluid shear contribution to the total friction is directly derived from the local velocity gradients and film thickness, requiring no empirical adjustment, while the solid–solid friction component is governed by the interfacial shear strength $\sigma_{\text{1}}$. A constant shear strength $\sigma_{\text{1}} = 0.1 \times E$ was adopted, resulting in excellent agreement with the experimental observations across mixed lubrication and full-film regimes. However, deviations emerge in the boundary regime, where the constant $\sigma_{\text{1}}$ assumption under-predicts friction. In this range, the real contact area remains nearly constant, and frictional variations are instead dominated by changes in molecular adhesion and interfacial shear rate. This indicates that $\sigma_{\text{1}}$ may depend on local kinematics and lubrication chemistry, as observed experimentally by \cite{vorvolakos2003effects}, who reported dependencies on sliding velocity, lubricant molecular weight and viscoelastic properties. Incorporating such rate-dependent interfacial strength could improve predictive accuracy in boundary lubrication conditions and could be informed by molecular dynamics or mesoscale simulations \cite{Vakis2018Sep}.

Nonetheless, the strong agreement between numerical predictions and experimental measurements across multiple lubrication regimes validates the robustness of the developed FSI framework. The model robustly captures the complex interplay between roughness-induced micro-contacts, elastic deformation and fluid film formation, demonstrating its ability to predict tribological performance without empirical calibration. From a practical perspective, most windscreen wiper systems operate within the mixed and full-film regimes, where a constant shear strength assumption remains a reasonable and industrially relevant approximation. While demonstrated here for a general soft contact geometry representative of systems such as lip seals and wiper blades, the framework provides a generalised platform applicable to diverse compliant systems such as biomedical elastomeric interfaces and soft robotic components. In each case, the predictive treatment of coupled roughness, fluid flow and elastic deformation offers a pathway to design more efficient and durable soft-contact technologies.

Future developments of the present framework can substantially broaden its scope, accuracy and cross-domain applicability. The authors are currently advancing the simulation framework to incorporate non-linear and viscoelastic constitutive models, enabling more realistic predictions of soft polymers, gels and biological tissues where time-dependent deformation plays a paramount role in lubrication performance \cite{Carmine2017, Afferrante2023}. Extending the formulation to three-dimensional anisotropic roughness and thermal coupling would as well enhance its relevance for dynamic, temperature-dependent interfaces in mechanical seals and biomedical devices. In addition to these extensions, the simulation framework can be revamped to capture further phenomena relevant to soft, rough contact systems. Integrating non-Newtonian lubricant rheology would expand its deployment for assessing smart fluids and bio-lubricants \cite{Attia2020}. The inclusion of poroelastic or swelling substrates, particularly in hydrogels and tissues, can facilitate the study of coupled fluid infiltration and solid deformation \cite{Reale2017, Holzapfel2025}. Accounting for adhesion, surface chemistry and capillary effects will enable the modelling of other interfacial phenomena such as wetting hysteresis under partially flooded conditions \cite{Liliane2008, Guo2017}, in addition to incorporating multiphase lubrication and wear-evolution models to assess the long-term durability and performance of sliding interfacial systems \cite{Tobias2021}. From a computational perspective, implementing adaptive coupling and convergence acceleration schemes, such as Aitken’s dynamic relaxation \cite{Ardah2023Jan}, could significantly improve numerical stability and computational efficiency. Moreover, replacing the simplified Gümbel cavitation model with a mass-conserving Elrod–Adams p–$\theta$ model can enable a more physically consistent representation of cavitation dynamics and mass transport processes across the interface \cite{Ardah2023Oct}. Furthermore, integrating molecular-scale and data-driven components, such as molecular dynamics-informed boundary conditions and machine-learning-based surrogate models \cite{Milanese2019, Holey2025, Kaliafetis2025}, could enable real-time digital twins for system optimisation, condition monitoring and uncertainty quantification.

\section{Conclusions}
\label{sec:Conclusions}
In this study, we have presented a modular and computationally efficient fluid-solid interaction (FSI) modelling framework for simulating the interfacial behaviour of soft contact systems across the full spectrum of lubrication regimes, from boundary lubrication to full-film lubrication. The framework couples the governing equations of hydrodynamic flow, rough surface contact mechanics and elastic deformation within a unified iterative model capable of resolving the non-linear interplay between fluid entrainment, asperity contact and material compliance.

Two complementary strategies were explored for modelling surface roughness. Persson’s statistical theory which provides analytical insight but relies on idealised fractal assumptions, in addition to a deterministic Conjugate Gradient Fast Fourier Transform (CG-FFT) approach that directly incorporates measured topographies. The deterministic method enables the derivation of load–separation and real contact area relationships through BEM-based simulations, which are then embedded as interpolation functions within the FSI framework. To address finite geometries and large deformations in compliant materials, a Reduced Stiffness Method (RSM) approach was developed by combining finite element modelling and model reduction to efficiently construct an accurate stiffness representation. Comparative analyses demonstrate that the RSM overcomes the limitations of traditional half-space approximations, achieving superior accuracy in predicting pressure distributions, film thickness and frictional response under high load and constrained configurations. The complete multiscale FSI model, integrating roughness, hydrodynamics and finite-body elasticity, was validated against experimental data from a wiper-blade-inspired contact system. Using realistic surface measurements and material parameters, the simulations accurately reproduced the measured Stribeck curves and lubrication transitions. This agreement highlights the robustness of the developed framework and its potential to serve as a versatile toolkit for the design, optimisation and performance evaluation of compliant lubricated interfaces.

\section{Acknowledgments}
Q. Wang and D. Dini would like to acknowledge the funding, materials and technical support received by Robert Bosch Produktie N.V. S. Ardah and D. Dini would like to acknowledge the support received from the Engineering and Physical Sciences Research Council, United Kingdom (EPSRC) via the InFUSE Prosperity Partnership EP/V038044/1. D. Dini would also like to acknowledge the funding received from the Engineering and Physical Sciences Research Council, United Kingdom (EPSRC) via grants EP/N025954/1, as well as the support provided by the Shell/RAEng Research Chair in Complex Engineering Interfaces (RCSRF2122-14-143).

\section{Availability of Data and Materials}
The data that support the findings of this study are available from the corresponding author upon reasonable request or at tribology@imperial.ac.uk.

\section{Conflict of Interest}
The authors declare that they have no known competing financial interests or personal relationships that could have appeared to influence the work reported in this paper.

\newpage
\appendix
\section{Conjugate Gradient Algorithm}
\label{app:CGFFT}
A detailed algorithmic implementation of the Conjugate Gradient (CG) method, adapted for solving large-scale rough surface contact problems, is presented below. The method is coupled with the Discrete Convolution and Fast Fourier Transform (DC-FFT) approach in order to accelerate numerical iterative convergence and simultaneously reduce the computational cost. This hybrid Conjugate Gradient Fast Fourier Transform (CG-FFT) formulation enables efficient evaluation of surface deformations through convolution operations in the Fourier domain. The iterative procedure unfolds as follows:
\begin{enumerate}
    \item \textbf{Initialisation:} Initialise pressure $p_{ij}$, contact set $I_c$, applied load $W$ and influence matrix matrix $D^{kl}_{ij}$.

    \item \textbf{Update gap:} Compute deformation $w_{ij}$ and update the interfacial separation:
    \begin{align}
        u_{ij} &\leftarrow w_{ij} + u_{ij}, \\
        u_{ij} &\leftarrow u_{ij} - \frac{1}{N_c} \sum_{(k,l) \in I_c} u_{kl},
    \end{align}
    where the nodal deformation $w_{ij}$ is computed using the DC-FFT method, and $N_c$ denotes the total number of nodes within the contact region $I_c$, defined as the set of indices $(i,j)$ for which the local pressure satisfies $p_{ij} > 0$.

    \item \textbf{Define auxiliary variable:} For the new gap $u_{ij}$, a variable $G$ is introduced as follows:
    \begin{equation}
        G = \sum_{(i,j) \in I_c} u_{ij}.
    \end{equation}

    \item \textbf{Conjugate direction update:} The conjugate direction $t$ at nodal position ($i,j$) can subsequently be obtained via the following relationships:
    \begin{equation}
        t_{ij} =
        \begin{cases}
        u_{ij} + \delta \left( \dfrac{G}{G_{\text{old}}} \right) t_{ij} & \text{if } (i,j) \in I_c \\
        0 & \text{if } (i,j) \notin I_c,
        \end{cases}
    \end{equation}
    where $\delta$ here is initialised as 0. Note that if $\delta=0$, $t_{ij}$ will coincide with the steepest
    decent direction.

    \item \textbf{Convolution and step size:} The convolution of $D^{kl}_{ij}$ and $t_{ij}$ is computed and adjusted by its mean value as follows:
    \begin{align}
        r_{ij} &= \sum_{(k,l) \in I_g} D^{kl}_{ij} t_{kl}, \\
        r_{ij} &\leftarrow r_{ij} - \frac{1}{N_c} \sum_{(k,l) \in I_c} r_{kl},
    \end{align}
    which is subsequently used to calculate the step size in the conjugate direction as follows:
    \begin{equation}
        \tau = \frac{\mathlarger{\mathlarger{\sum}}_{(i,j) \in I_c} u_{ij} t_{ij}}{\mathlarger{\mathlarger{\sum}}_{(i,j) \in I_c} r_{ij} t_{ij}}.
    \end{equation}

    \item \textbf{Pressure update:} After storing the current pressure as $p^{\text{old}} = p$, the contact pressure is updated according to the following relationship:
    \begin{equation}
        p_{ij} \leftarrow p_{ij} - \tau \, t_{ij} \quad \text{for } (i,j) \in I_c.
    \end{equation}

    \item \textbf{Inequality enforcement:} To enforce the inequalities, all negative pressure values are initially set to zero. Subsequently, nodes exhibiting zero pressure in conjunction with negative interfacial separation, indicative of physical overlap, are identified as requiring correction. These nodal points define the overlapping domain, where contact is present but not yet accounted for in the pressure solution, and thus the local pressure must be reintroduced to restore mechanical admissibility. The overlapping domain is determined as follows:
    \begin{align}
        I_{\text{ol}} &= \left\{ (i,j) \in I_g \mid p_{ij} = 0, \; u_{ij} < 0 \right\}.
    \end{align}
    Therefore, the following can be implied:
    \begin{align}
        \begin{cases}
            \delta &= 1 \quad \quad \text{if} \quad I_{\text{ol}} = \emptyset, \\
            \delta &= 0 \quad \quad \text{otherwise}.
        \end{cases}      
    \end{align}
    Hence, the pressure in the new contact area is adjusted using the following:
    \begin{align}
        p_{ij} &\leftarrow p_{ij} - \tau u_{ij} \quad \text{for } (i,j) \in I_{\text{ol}}.
    \end{align}

    \item \textbf{Load balance:} Load balance is finally performed, and the new pressure is adjusted based on the current load $F_n$ and the applied load $W$ as follows:
    \begin{align}
        F_n &= \Delta x \Delta y \sum_{(i,j) \in I_g} p_{ij}, \\
        p_{ij} &\leftarrow p_{ij} \left(\frac{F_n}{W}\right) \quad \quad \text{for } (i,j) \in I_g,
    \end{align}

    \item \textbf{Convergence check:} Convergence check for pressure to exit the iteration uses the following criterion:
    \begin{equation}
        \frac{\Delta x \Delta y}{W} { \mathlarger{\mathlarger{\sum}}_{(i,j) \in I_g} \left| p_{ij} - p^{\text{old}}_{ij} \right|} \leq 10^{-10}.
    \end{equation}
\end{enumerate}

\newpage
\section{Mixed–Elastohydrodynamic Lubrication (MEHL) Algorithm}
\label{app:MEHL}
The Mixed–Elastohydrodynamic Lubrication (MEHL) solver developed in this work combines multiple numerical strategies to capture the coupled interactions between fluid pressure, solid contact, elastic deformation and interfacial separation. The method integrates a Reynolds Equation Sweep (RES), a Fredholm Solver (FS) and a Central Separation Updater (CSU), supported by the Reduced Stiffness Method (RSM) for efficient deformation calculations in compliant systems. The computational procedure proceeds through the following key steps:

\begin{enumerate}
\item \textbf{Initialisation:} Initialise pressure $p_f$, solid pressure $p_s$, interfacial separation $u$ and central separation $u_c$ using results from a previous lower-speed simulation for enhanced convergence.

\item \textbf{Fluid pressure update (RES):} Update fluid pressure $p_f$ by performing a one-pass Gauss–Seidel sweep of the Reynolds equation as follows:
\begin{equation}
        \begin{split}
            L_i^m =\ & {u_{0}}_{i+1} - {u_{0}}_{i-1} + \mathlarger{\sum}_j \bigg( K_{i+1,j} - K_{i-1,j} \bigg) \bigg( [p_f]^m + [p_s]^m \bigg)_j \\
            & - \frac{1}{\Delta x_i} \Bigg\{ 
                \bigg( \varepsilon_{i-1/2} \bigg) \bigg[ p_f \bigg]^m_{i-1} 
                + \bigg( \varepsilon_{i+1/2} \bigg) \bigg[ p_f \bigg]^m_{i+1}  + \bigg( \varepsilon_{i-1/2} + \varepsilon_{i+1/2} \bigg) \bigg[ p_f \bigg]^m_i
            \Bigg\},
        \end{split}
\end{equation}
where $\varepsilon = u^3/6\nu\eta$. As a result, and based on the Newton-Raphson method, the fluid pressure is updated as follows:
\begin{equation}
    \mathlarger{\mathlarger{\sum}}_C \left[ \frac{\partial L_i}{\partial \left[ p_f \right]_C} \right]_{m-1} \bigg( [p_f]^{m} - [p_f]^{m-1} \bigg)_C = -L^{m-1}_{i}.
\end{equation}
To account for cavitation effects, the overall contact domain ($C$) is partitioned into three computational zones: the high-pressure region ($C_1$), the low-pressure region ($C_2$), and the cavitation region ($C_3$). Within each of these zones, the Newton–Raphson method is reformulated accordingly to accommodate the local pressure characteristics as follows:
\begin{equation}
        \begin{aligned}
            \text{In } C1: \quad
            - L_i^{m-1} = \; &
            \bigg[ \frac{\partial L_i}{\partial [p_f]_{i-2}} \bigg]^{m-1} \bigg( [p_f]^m - [p_f]^{m-1} \bigg)_{i-2} 
            + \bigg[ \frac{\partial L_i}{\partial [p_f]_{i-1}} \bigg]^{m-1} \bigg( [p_f]^m - [p_f]^{m-1} \bigg)_{i-1} \\[8pt] 
            & + \bigg[ \frac{\partial L_i}{\partial [p_f]_{i+1}} \bigg]^{m-1} \bigg( [p_f]^m - [p_f]^{m-1} \bigg)_{i+1} 
            + \bigg[ \frac{\partial L_i}{\partial [p_f]_{i+2}} \bigg]^{m-1} \bigg( [p_f]^m - [p_f]^{m-1} \bigg)_{i+2}  \\[8pt] 
            & + \bigg[ \frac{\partial L_i}{\partial [p_f]_{i}}   \bigg]^{m-1} \bigg( [p_f]^m - [p_f]^{m-1} \bigg)_{i} ,
        \end{aligned}
\end{equation}

\vspace{1em}

\begin{equation}
        \begin{aligned}
            \text{In } C2: \quad
            - L_i^{m-1} = \; &
            \bigg[ \frac{\partial L_i}{\partial [p_f]_{i-1}} \bigg]^{m-1} \bigg( [p_f]^m - [p_f]^{m-1} \bigg)_{i-1} \\[8pt]
            & + \bigg[ \frac{\partial L_i}{\partial [p_f]_{i+1}} \bigg]^{m-1} \bigg( [p_f]^m - [p_f]^{m-1} \bigg)_{i+1} \\[8pt]
            & + \bigg[ \frac{\partial L_i}{\partial [p_f]_{i}} \bigg]^{m-1} \bigg( [p_f]^m - [p_f]^{m-1} \bigg)_{i},
        \end{aligned}
\end{equation}

\vspace{1em}

\begin{align}
        \text{In } C3: \quad
        - L_i^{m-1} = \; 
        \bigg[ \frac{\partial L_i}{\partial [p_f]_{i}} \bigg]^{m-1} \bigg( [p_f]^m - [p_f]^{m-1} \bigg)_{i}.
\end{align}

\item \textbf{Central Separation Update (CSU):} A convergence check is subsequently performed prior to relaxing the fluid pressure using a relaxation factor $\alpha_f$. Following this, the central separation is updated within the Central Separation Updater (CSU) based on the increment $\Delta u_c$, which is computed as follows:
\begin{equation}
    \Delta u_c^m =
    \begin{cases}
        - \alpha_c \Delta F^m & \text{if } \Delta F_c^m \cdot \Delta F^m \leq 0 \\[1ex]
        0 & \text{if } \Delta F_c^m \cdot \Delta F^m > 0,
    \end{cases}
\end{equation}
where
\begin{equation}
    \Delta F^m = \left[ W - \Delta x \Delta y \mathlarger{\mathlarger{\sum}}_i \bigg( [p_f]^m_i + [p_s]^m_i \bigg) \right],
\end{equation}

\begin{equation}
    \Delta F_c^m = \mathlarger{\mathlarger{\sum}}_i \bigg( [p_f]^m_i + [p_s]^m_i \bigg) + \mathlarger{\mathlarger{\sum}}_i \bigg( [p_f]^{m-1}_i + [p_s]^{m-1}_i \bigg).
\end{equation}

\item \textbf{Reduced stiffness matrix transformation:} Replace traditional influence matrix with the equivalent stiffness matrix $\big[\mathbf{EK}\big]$ derived from the reduced stiffness matrix $\big[\mathbf{K}_r\big]$ as follows:
\begin{equation}
    \big[ \mathbf{EK} \big]_{ij} = \frac{1}{2} \left( \Delta s_{j1} + \Delta s_{j2} \right)
    \left( \big[ \mathbf{K}_r \big]_{2i,2j-1} \cos \theta_j + \big[ \mathbf{K}_r \big]_{2i,2j} \sin \theta_j \right),
\end{equation}
where $\Delta s$ and $\theta$ denote the spacing and angle between adjacent nodes.

\item \textbf{Separation and solid pressure update (FS):} Given the fluid pressure and the central separation, a Fredholm-type residual equation is defined as follows:
\begin{equation}
    g_i = -1 + \frac{u_c {u_{0}}_i + \mathlarger\sum_j \big[ \mathbf{EK} \big]_{ij} \bigg[ p_f + p_s \bigg]_j}{u_i}.
\end{equation}
Thus, and using a relaxation factor $\alpha_{fs}$, the new separation profile is obtained as follows:
\begin{equation}
    u_i^{l} = u_i^{l-1} + \left[ 1 + \alpha_{fs} \frac{g_{i}^{l-1}}{1 + g_{i}^{l-1} - \mathlarger\sum_j \big[ \mathbf{EK} \big]_{ij} \left( \dfrac{\partial p_s}{\partial u_j} \right)} \right].
    \label{eq:MEHL_separation_update}
\end{equation}

\item \textbf{Convergence check:} Evaluate residuals for pressure and separation. If both meet the specified tolerance criteria, the MEHL module terminates. Otherwise, return to Step 2.
\end{enumerate}

\newpage
\section{Benchmarking the Reduced Stiffness Matrix Method}
\label{app:FFT_IC}
To further assess the robustness and computational reliability of the proposed framework, a benchmark comparison is conducted between the Reduced Stiffness Matrix (RSM), denoted as $\bigl[ \mathbf{K}_\mathrm{r} \bigr]$, and the conventional influence coefficient matrix $\bigl[ \mathbf{I} \bigr]$. While the main body of the paper demonstrated the accuracy of the RSM in capturing the frictional behaviour of soft, lubricated contacts, it is essential to evaluate its performance relative to the classical formulation, particularly under higher loading conditions where boundary effects become significant. Comparative simulations were therefore carried out using the same contact configuration described in \autoref{subsec:Exp_Validation}, but under an increased applied load of 60 N/m. Surface roughness parameters and the contact pressure approximation were adopted from \cite{persson2009transition} to maintain consistency with prior models while isolating the impact of the deformation representation. This ensures that any observed differences stem solely from the distinct elastic formulations, rather than from topographical or rheological variations.

\begin{figure}[H]
        \centering
        \includegraphics[width=\linewidth]{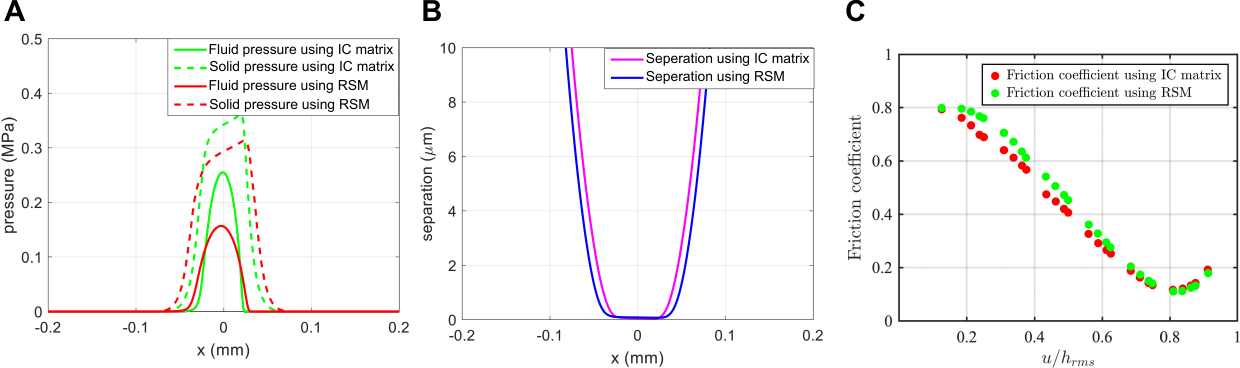}
        \caption{Comparative performance of the Reduced Stiffness Matrix (RSM) and the conventional influence coefficient matrix approaches for a soft lubricated contact operating under a sliding velocity of $\mathrm{5} \times 10^{-4}$ m/s and an applied load of 60 N/m. (\textbf{A}) Fluid and solid pressure distributions along the sliding (entraining) direction. (\textbf{B}) Interfacial separation profiles, representing the local lubricant film thickness. (\textbf{C}) Predicted coefficient of friction (Stribeck) curves spanning the full range of lubrication regimes, from boundary to elastohydrodynamic lubrication.}
        \label{fig:RSM_IC}
\end{figure}

As shown in \autoref{fig:RSM_IC}, the RSM produces slightly larger film thicknesses and a broader real contact area, accompanied by lower peak solid pressures; an effect attributed to the finite-body compliance and boundary constraints explicitly incorporated into the RSM formulation. Despite these localised variations, the macroscopic tribological response, represented by the Stribeck curve in \autoref{fig:RSM_IC}\textbf{C}, remains nearly identical between the two approaches across the mixed lubrication regime. This confirms that both models yield consistent global behaviour under moderate deformation, though the RSM provides enhanced physical fidelity near the contact edges. It should be emphasised that the current analysis focuses on linear elastic materials with small geometric deformations. In realistic systems—, such as soft polymers, biomaterials and adaptive elastomeric components, both material nonlinearity and large-strain effects become non-negligible. The modular formulation of the RSM makes it readily extendable to such cases, offering a computationally efficient yet physically rigorous alternative to full finite element coupling. As such, the RSM provides a scalable foundation for next-generation FSI solvers capable of addressing complex, non-linear deformation and heterogeneous geometries.

\newpage
\bibliography{mybibfile}

\end{document}